\documentclass[aps,prx,reprint,groupedaddress, showpacs]{revtex4-1}
\usepackage{float}
\usepackage{pdfpages}

\usepackage{feynmp}
\usepackage{enumitem}
\usepackage{amssymb}
\usepackage{amsmath}
\usepackage{mathtools}
\usepackage{MnSymbol}
\usepackage{esvect}
\usepackage{color}
\usepackage{bm}

\newcommand{\true}{{\rm true}}

\newcommand{\row}{{\rm row}}

\newcommand{\bea}{\begin{eqnarray}}
\newcommand{\eea}{\end{eqnarray}}
\newcommand{\beas}{\begin{eqnarray*}}
\newcommand{\eeas}{\end{eqnarray*}}
\newcommand{\eq}[1]{Eq.\ (\ref{#1})}

\newcommand{\data}{{\rm data}}
\newcommand{\set}[1]{\left\{ #1 \right\}}

\newcommand{\bra}[1]{\left\langle #1 \right|}

\newcommand{\braket}[1]{\left\langle #1 \right\rangle}

\definecolor{red}{rgb}{1,0,0}

\renewcommand{\vec}[1]{{\boldsymbol{#1}}}
\newcommand{\mat}[2]{\left( \begin{array}{#1} #2 \end{array} \right)}

\begin{document}

%\preprint{}

\title{Unification of field theory and maximum entropy methods for\\  learning probability densities}
%\title{Field theory, maximum entropy, and density estimation}
%\title{Maximum entropy density estimation and\\ Bayesian field theory in the infinite smoothness limit.}


\author{Justin B. Kinney}
\email[Email correspondence to ]{jkinney@cshl.edu}
%\thanks{
\affiliation{Simons Center for Quantitative Biology, Cold Spring Harbor Laboratory, Cold Spring Harbor, New York 11724, USA}

%\date{\today}

% 207 words; I LIKE THIS ABSTRACT
\begin{abstract}
The need to estimate smooth probability distributions (a.k.a.\ probability densities) from finite sampled data is ubiquitous in science. Many approaches to this problem have been described, but none is yet regarded as providing a definitive solution. Maximum entropy estimation and Bayesian field theory are two such approaches. Both have origins in statistical physics, but the relationship between them has remained unclear.  Here I unify these two methods by showing that every maximum entropy density estimate can be recovered in the infinite smoothness limit of an appropriate Bayesian field theory. I also show that Bayesian field theory estimation can be performed without imposing any boundary conditions on candidate densities, and that the infinite smoothness limit of these theories recovers the most common types of maximum entropy estimates. Bayesian field theory thus  provides a natural test of the maximum entropy null hypothesis and, furthermore, returns an alternative (lower entropy) density estimate when the maximum entropy hypothesis is falsified. The computations necessary for this approach can be performed rapidly for one-dimensional data, and software for doing this is provided. 
\end{abstract}

% 02.50.-r		Probability theory, stochastic processes, and statistics
% 89.70.Cf		Entropy in information theory
% 02.60.-x 		Numerical approximation and analysis
% 11.10.Lm		Nonlinear (nonlocal) field theory


%\keywords{}

\maketitle

\section{Introduction}

% Motivate the density estimation problem 
Research in nearly all fields of science routinely calls for the estimation of smooth probability densities from finite sampled data \cite{Silverman:1986, Scott:1992}. Indeed, the presence of histograms in a large fraction of the scientific literature attests to this need. But the problem of how to go beyond a histogram and recover a smooth probability distribution has yet to find a definitive solution, even in one dimension. 

% Give background on common methods
The reader might find this state of affairs surprising. Many different methods for estimating smooth probability densities are well known and commonly used.  One of the most popular methods is kernel density estimation \cite{Silverman:1986, Scott:1992}. Kernel density estimation is easy to carry out, but this approach has little theoretical justification and there is no consensus on certain basic aspects of its use, such as how to choose a kernel width or how to treat data points near a boundary \cite{Eggermont:2001}. Bayesian inference of a Gaussian mixture model is another common method. This approach, however, requires that one assume an explicit functional form of the density that one wishes to learn. 

% Give background on statistical physics methods. 
Concepts from statistical physics have given rise to two alternative approaches to the density estimation problem: maximum entropy (MaxEnt) \cite{Jaynes:1957, Mead:1984} and Bayesian field theory \cite{Bialek:1996, Nemenman:2002, Kinney:2014, Ensslin:2009, Lemm:2003, Holy:1997, Periwal:1997, Aida:1999, Schmidt:2000}. Each of these approaches has a firm but distinct theoretical basis. MaxEnt derives from the principle of maximum entropy as described by Jaynes in 1957 \cite{Jaynes:1957}. Bayesian field theory, which is also referred to as ``information field theory'' in some of the literature \cite{Ensslin:2009}, instead uses the standard methods of Bayesian inference together with priors that weight possible densities according to an explicit measure of smoothness without requiring a particular functional form.  Perhaps because the principles underlying these two methods are different, the relationship between these approaches has remained unclear. 

% Explain maximum entropy estimation OK
MaxEnt density estimation is carried out as follows. One first uses sampled data to estimate values for a chosen set of moments, e.g., mean and variance. Typically, all moments up to some specified order are selected \cite{Mead:1984, Ormoneit:1999}. The probability density that matches these moments while having the maximum possible entropy is then adopted as one's estimate. All other information in the data is discarded. One can therefore think of the MaxEnt estimate as a null hypothesis reflecting the assumption that there is no useful information in the data beyond the values of the specified moments \cite{Good:1963}. 

% Describe how field theoretic inference works OK
In the Bayesian field theory approach, one first defines a prior on the space of continuous probability densities. This prior is formulated using a scalar field theory that favors smooth probability densities over rugged ones. The data are then used to compute a Bayesian posterior, and from this one identifies the maximum \emph{a posteriori} (MAP) density estimate. Simple field theory priors require that one assume an explicit smoothness length scale $\ell$. However, an optimal value for $\ell$ can be learned from the data in a natural way if one instead adopts a prior formed from a scale-free mixture of these simple field theories \cite{Bialek:1996, Nemenman:2002, Kinney:2014}. Scale-free Bayesian field theories thus provide a way to estimate probability densities without having to specify any tunable parameters. 

% Describe the advantages of field theoretic density estimation
One problem with the field theory priors that have been considered for this purpose thus far \cite{Bialek:1996, Nemenman:2002, Kinney:2014} is that they impose boundary conditions on candidate densities. This assumption of boundary conditions is standard practice in physics; it greatly aids analytic calculations and is often well-motivated by physical reasoning. In the density estimation context, however, such boundary conditions limit the types of data sets for which such field theory priors would be appropriate. MaxEnt, by contrast, does not impose any boundary conditions on the density estimates it provides. 

% Explain result of removing boundary conditions
Here I describe a class of Bayesian field theory priors that have no boundary conditions. These priors yield MAP density estimates that exactly match the first few moments of the data. In the $\ell \to \infty$ limit, such MAP estimates become identical to MaxEnt estimates constrained by these same moments. More generally, I show that a MaxEnt density estimate matched to any moments of the data can be recovered from Bayesian field theory in the infinite smoothness limit; one need only choose a field theory prior that defines ``smoothness'' appropriately. 

% Explain test of MaxEnt
This unification of Bayesian field theory and MaxEnt density estimation further suggests a natural way to test the validity of the MaxEnt hypothesis against one's data. If Bayesian field theory identifies $\ell = \infty$ as being optimal for one's data set, the MaxEnt hypothesis is validated. If instead the optimal $\ell$ is finite, the MaxEnt hypothesis is rejected in favor of a nonparametric density estimate that matches the same moments of the data but has lower entropy. 

% Describe mathematical parts of paper
This paper is structured as follows. Section II describes the derivation of an action, $S_\ell$, that governs the posterior probability of densities under a specific class of Bayesian field theories. Section III describes how the MAP density, which minimizes this action, can be uniquely derived without assuming any boundary conditions. A differential operator I call the ``bilateral Laplacian'' plays a central role in eliminating these boundary conditions. 

Section IV shows that such MAP density estimates reduce to MaxEnt estimates in the $\ell \to \infty$ limit. Section V derives an expression for a quantity, the ``evidence ratio'' $E(\ell)$, that allows one to select the optimal value for $\ell$ given the data. The large $\ell$ behavior of this evidence ratio is shown to be characterized by a ``$K$ coefficient,'' the sign of which provides a novel analytic test of the MaxEnt assumption. 

% Describe computational parts of paper
Section VI formalizes a discrete-space representation of this Bayesian field theory inference procedure. In addition to being essential for the computational implementation of this method, this discrete representation greatly clarifies why no boundary conditions are required to derive the MAP density when one makes use of the bilateral Laplacian. Section VII describes how to compute the MAP density (to a specified precision) at all length scales $\ell$. Section VIII illustrates this density estimation approach on simulated data sets. A summary and discussion are provided in section IX. 

% Describe appendices
Detailed derivations of various results from sections II through VI are provided in Appendices A-D. Appendix E presents details of a predictor-corrector homotopy algorithm that allows the density estimation computations described in this paper to be carried out. An open source software implementation of this algorithm for one-dimensional density estimation is provided \footnote{Available at \texttt{https://github.com/jbkinney/14{\textunderscore}maxent}}. Finally, Appendix F gives an expanded discussion of how Bayesian field theory relates to earlier work in statistics on ``maximum penalized likelihood'' \cite{Silverman:1982, Eggermont:2001, Gu:2013}. 

%
% Bayesian field theory
%
\section{Bayesian field theory}

% Set up problem in 1D and notation
The main results of this paper are elaborated in the context of one-dimensional density estimation.  Many of our results are readily extended to higher dimensions, however, at least in principle. This issue is discussed in more detail later on. 

% Explain method described in my previous paper
Suppose we are given $N$ data points $x_1, x_2, \ldots, x_N$ sampled from a smooth probability density $Q_\true(x)$ that is confined to an interval of length $L$. Our goal is to estimate $Q_\true$ from these data.  Following \cite{Kinney:2014}, we first represent each candidate density $Q(x)$ in terms of a real field $\phi(x)$ via 
\bea
Q(x) &=& \frac{e^{-\phi(x)}}{\int dx' e^{-\phi(x')}}. \label{eq:Q_phi_map}
\eea
This parametrization ensures that $Q$ is positive and normalized \footnote{Integrals over $x$ are restricted to the interval of length $L$.}. Next we adopt a field theory prior on $\phi$. Specifically we consider priors of the form
\bea
p(\phi | \ell) = \frac{e^{-S^0_\ell[\phi]}}{Z^0_\ell}
\eea
where 
\bea
S^0_\ell[\phi]  = \int \frac{dx}{L} \frac{\ell^{2 \alpha}}{2} (\partial^\alpha \phi)^2,~~~~\label{eq:prior}
\eea
is the ``action'' corresponding to this prior and
\bea
Z^0_\ell = \int \mathcal{D}\phi\,e^{-S^0_\ell[\phi]}
\eea
is the associated partition function. The real parameter $\ell$ is a length scale below which fluctuations in $\phi$ are strongly damped. The parameter $\alpha$, on the other hand, reflects a fundamental choice in how we define ``smoothness.'' In this paper we consider arbitrary positive integer values of $\alpha$, for reasons that will become clear. Note, however, that previous work has explored the consequences of using non-integer values of $\alpha$ \cite{Nemenman:2002}.

As shown in Appendix A, this choice of prior allows us to compute an exact posterior probability $p(\phi | \data, \ell)$ over candidate densities. We find that
\bea
p(\phi | \data, \ell) &=& \frac{e^{- S_\ell [\phi]} }{Z_\ell}, \label{eq:posterior}
\eea
where
\bea
S_\ell[\phi] &=& \int \frac{dx}{L} \left\{ \frac{\ell^{2 \alpha}}{2} (\partial^\alpha \phi)^2 + N L R \phi + N e^{-\phi} \right\} \label{eq:action}
\eea
is a nonlinear action, 
\bea
Z_\ell = \int \mathcal{D}\phi\,  e^{- S_\ell [\phi]} \label{eq:posterior_partition_function}
\eea is the corresponding partition function, and 
\bea
R(x) = N^{-1} \sum_{n=1}^N \delta(x - x_n)
\eea
is the raw data density. 

The derivation of \eq{eq:action} is somewhat subtle. In particular, the action $S_\ell[\phi]$ gives a posterior probability $p(\phi | \data, \ell)$ that is not related to $p(\phi | \ell)$ via Bayes's rule. However, upon marginalizing over the constant component of $\phi$ one finds that $p(Q | \data, \ell)$ is indeed related to $p(Q | \ell)$ via Bayes's rule. This latter fact is sufficient to justify the use of \eq{eq:action} in what follows. See Appendix A for details. 

%
% Boundary conditions are unnecessary
%
\section{Eliminating boundary conditions}

% Describe how boundary conditions were used to get the EOM, and what the problems with boundary conditions is
The MAP field $\phi_\ell$ is defined as the field that minimizes the action $S_\ell$. To obtain a differential equation for $\phi_\ell$, previous work \cite{Bialek:1996, Nemenman:2002, Kinney:2014} imposed periodic boundary conditions on $\phi$ and used integration by parts to derive
\bea
\ell^{2 \alpha} (-1)^{\alpha} \partial^{2 \alpha} \phi_\ell + N L R - N e^{-\phi_\ell} = 0. \label{eq:classic_eom}
\eea
With the periodic boundary conditions in place, this differential equation has a unique solution. However, imposing these boundary conditions amounts to assuming that $Q_\true(x)$ is the same at both ends of the $x$-interval. It is not hard to imagine data sets for which this assumption would be problematic. 

% Explain that this does not prove boundary conditions are needed
It is true, of course, that \eq{eq:classic_eom} requires boundary conditions in order to have a unique solution. The reason boundary conditions are needed is the appearance of the of the standard $\alpha$-order Laplacian operator, $(-1)^\alpha \partial^{2 \alpha}$.  However, we assumed boundary conditions on $\phi$ in order to derive \eq{eq:classic_eom} in the first place.  It therefore has not been established that boundary conditions are required for $S_\ell[\phi]$ to have a unique minimum. 

% Explain boundary conditions are unnecessary
In fact, $S_\ell[\phi]$ has a unique minimum without the imposition of any boundary conditions on $\phi$.  The boundary conditions on $\phi$  assumed in previous work \cite{Bialek:1996, Nemenman:2002, Kinney:2014} are therefore unnecessary. Indeed, from \eq{eq:action} alone we can derive a differential equation that uniquely specifies the MAP field $\phi_\ell$. 

We start by rewriting the action as
\bea
S_\ell[\phi] &=& \int \frac{dx}{L} \left\{ \frac{\ell^{2 \alpha}}{2} \phi \Delta^\alpha \phi + N L R \phi + N e^{-\phi} \right\} \label{eq:action_rexpressed}
\eea
where the differential operator $\Delta^\alpha$ is defined by the requirement that
\bea
\varphi \Delta^\alpha \phi &=& (\partial^\alpha \varphi) (\partial^\alpha \phi) \label{eq:bl_def}
\eea 
for any two fields $\varphi$ and $\phi$. In what follows we refer to $\Delta^\alpha$ as the ``bilateral Laplacian of order $\alpha$.'' Note that $\Delta^\alpha$ is a positive semi-definite operator, since
\bea
\int dx\, \phi \Delta^\alpha \phi = \int dx\, (\partial^\alpha \phi)^2 \ge 0.
\eea
for every real field $\phi$. 

We now prove that $\phi_\ell$ is unique by showing that $S_\ell[\phi]$ is a strictly convex function of $\phi$ when $N > 0$. Consider the change in $S_\ell[\phi]$ upon the perturbation $\phi \to \phi + \epsilon \psi$, where $\phi$ and $\psi$ are two real fields and $\epsilon$ is an infinitesimal number and the field $\psi$ is normalized so that $L^{-1} \int dx\, \psi^2 = 1$. The action will change by an amount 
\bea
S_\ell[\phi + \epsilon \psi] &=&  S_\ell[\phi] + \epsilon \int dx\, \psi \left. \frac{\delta S}{\delta \phi} \right|_\phi  \label{eq:action_perturbation} \\
& & + \frac{\epsilon^2}{2} \int dx \left\{ \frac{\ell^{2 \alpha}}{L} \psi \Delta^\alpha \psi + \frac{N}{L} e^{-\phi} \psi^2 \right\} + \cdots~~~~ \nonumber
\eea
Because $\Delta^\alpha$ is positive semi-definite, the $O(\epsilon^2)$ term will be bounded from below by $ \epsilon^2 N \exp [ - \max(\phi) ]$ and must therefore be positive.  The Hessian of $S_\ell$ is therefore positive definite at every $\phi$, establishing the strict convexity of $S_\ell$ and thus the uniqueness of $\phi_\ell$.

The requirement that $\delta S_\ell / \delta \phi = 0$ gives the following differential equation for $\phi_\ell$:
\bea
0 = \ell^{2 \alpha} \Delta^\alpha \phi_\ell + N L R - N e^{-\phi_\ell}. \label{eq:eom}
\eea
From the argument above we see that this differential equation, unlike \eq{eq:classic_eom}, has a unique solution without the imposition of any boundary conditions on $\phi_\ell$. 

This lack of a need for boundary conditions in \eq{eq:eom}, despite the need for boundary conditions in \eq{eq:classic_eom}, is due to a fundamental difference between the standard Laplacian and the bilateral Laplacian. This difference occurs only at the boundaries of the $x$-interval. Roughly speaking, $\Delta^\alpha \phi$ is well-defined at both $x_\mathrm{min}$ and $x_\mathrm{max}$, whereas $(-1)^\alpha \partial^{2 \alpha} \phi$ is not. This point will be clarified in Section VI, when we formulate our Bayesian field theory approach on a finite set of grid points.

In the interior of the $x$-interval, however, the bilateral Laplacian is identical to the standard Laplacian. To see this, we integrate \eq{eq:bl_def} and use integration by parts to derive
\bea
\int dx\,\varphi \Delta^\alpha \phi &=& \int dx\, \varphi \left[(-1)^\alpha  \partial^{2 \alpha} \right]  \phi \label{eq:aa} \\
&& + \sum_{b=0}^{\alpha-1} \left[ (-1)^b (\partial^{\alpha -b-1} \varphi) (\partial^{\alpha+b} \phi) \right]_{x_\mathrm{min}}^{x_\mathrm{max}}. \nonumber
\eea
The second term on the right hand side vanishes if the test function $\varphi$ is chosen so that $\partial^b \varphi = 0$ at $x_\mathrm{min}$ and $x_\mathrm{max}$ for $b = 0, 1, \ldots, \alpha-1$. The value of such test functions $\phi$ within the interior of the interval are unconstrained, and so 
\bea
\Delta^\alpha \phi(x) = (-1)^\alpha \partial^{2 \alpha} \phi(x),~~\mathrm{for~all}~~~x_\mathrm{min} < x < x_\mathrm{max}.~~~~~~
\eea

%
% Connection to maximum entropy
%
\section{Connection to maximum entropy} 

From its definition in \eq{eq:bl_def}, we see that the bilateral Laplacian is symmetric and real. This operator is therefore Hermitian and possesses a complete set of orthonormal eigenvectors with corresponding real eigenvalues. See Appendix B for a discussion of the spectrum of the bilateral Laplacian. 

The kernel of $\Delta^\alpha$ is particularly relevant to the density estimation problem. A field $\phi$ is in the kernel of $\Delta^\alpha$ if and only if
\bea
\int dx\, \phi \Delta^\alpha \phi = \int dx\, (\partial^\alpha \phi)^2 = 0.
\eea
From this we see that the kernel of $\Delta^\alpha$ is equal to the space of polynomials of order $\alpha - 1$. 

In particular, $\phi = 1$ is in the kernel of $\Delta^\alpha$ for all positive integers $\alpha$. As a result, multiplying \eq{eq:eom} on the left by unity and integrating gives $\int dx\, e^{-\phi_\ell} = L$. The MAP density $Q_\ell$, which is defined in terms of $\phi_\ell$ by \eq{eq:Q_phi_map},  is thereby seen to have the simplified form,
\bea 
Q_\ell = \frac{e^{-\phi_\ell}}{L}. \label{eq:MAP_field_to_MAP_density}
\eea
If we multiply \eq{eq:eom} on the left by other polynomials of order $\alpha-1$ and integrate, we further find that 
\bea
\int dx\,Q_\ell\, x^k = \int dx\,R\,x^k,~~~k = 1, \ldots, \alpha-1.~~~ \label{eq:moment_matching}
\eea
Therefore, at every length scale $\ell$, the first $\alpha-1$ moments of the MAP density $Q_\ell$ exactly match those of the data.

% Describe connection to maximum entropy density
At $\ell = \infty$, the MAP field $\phi_\infty$ is restricted to the kernel of the bilateral Laplacian.  The corresponding density thus has the form
\bea
Q_\infty(x) &=& \frac{1}{L} \exp \left( - \sum_{k=0}^{\alpha-1} a_k x^k \right), \label{eq:connection_to_maxent}
\eea
where the values of the coefficients $a_k$ are determined by Eqs.\ \ref{eq:MAP_field_to_MAP_density} and \ref{eq:moment_matching}. $Q_\infty$ is therefore identical to the MaxEnt density that matches the first $\alpha-1$ moments of the data \cite{Mead:1984}. 

At $\ell = 0$, the kinetic term in \eq{eq:eom} vanishes. As a result, setting $\delta S_0 / \delta \phi = 0$ gives
\bea
Q_0(x) &=& R(x).
\eea
We therefore see that the MAP density $Q_0$ is simply the ``histogram'' of the data, i.e.\ the normalized sum of delta functions centered at each data point. When we formulate our inference procedure on a grid in section VI, we will see that $Q_0 = R$ indeed becomes a bona fide histogram with bins defined by our choice of grid.  

The set of MAP densities $Q_\ell$ thus forms a one-parameter ``MAP curve'' in the space of probability densities extending from the data histogram at $\ell = 0$ to the MaxEnt density at $\ell = \infty$. Every density $Q_\ell$ along this MAP curve exactly matches the first $\alpha - 1$ moments of the data. 

More generally, the MaxEnt density estimate constrained to match any set of moments can be recovered in the infinite smoothness limit of an appropriate Bayesian field theory. To see this, consider a MaxEnt estimate $Q_\mathrm{ME}$ chosen to satisfy the generalized moment-matching criteria
\bea
\int dx\, Q_\mathrm{ME}\, f_j = \int dx\, R\, f_j,~~~j=1,2,\ldots,J
\eea
for some set of user-specified functions $f_1(x)$, $f_2(x)$, $\ldots$, $f_J(x)$. A Bayesian field theory that recovers this MaxEnt estimate in the infinite smoothness limit can be readily constructed by using a prior defined by the action
\bea
S_\xi^0[\phi] = \int \frac{dx}{L} \frac{\xi}{2} \phi \Delta \phi \label{eq:general_prior_action}
\eea
where $\xi$ is the (positive) smoothness parameter and $\Delta$ is a positive semidefinite operator whose kernel is spanned by the specified functions $f_1$, $f_2$, $\ldots$, $f_J$ together with the constant function $f_0(x) = 1$. The posterior probability on $\phi$ will then be governed by the action
\bea
S_\xi[\phi] &=& \int \frac{dx}{L} \left\{ \frac{\xi}{2} \phi \Delta^\alpha \phi + N L R \phi + N e^{-\phi} \right\} \label{eq:general_posterior_action}.
\eea
Following the same line of reasoning as above, we find that the MAP density $Q_\xi$, corresponding to the field $\phi_\xi$ that minimizes $S_\xi$, will satisfy
\bea
\int dx\, Q_\xi\, f_j = \int dx\, R\, f_j,~~~j = 0, 1, \ldots, J\label{eq:general_contraints}
\eea
regardless of the value of $\xi$. In the infinite smoothness limit ($\xi \to \infty$), the MaxEnt density will be recovered, i.e.
\bea
Q_\infty(x) = \frac{1}{L} \exp \left( - \sum_{j=0}^J a_j f_j(x) \right) = Q_{ME}(x)
\eea
where the coefficients $a_0, a_1, \ldots, a_j$ are determined by the constraints in \eq{eq:general_contraints}.

%
% Describe Laplace approximation of the evidence
%
\section{Choosing the length scale} 

To determine the optimal value for $\ell$, we compute $p(\data | \ell) =  \int \mathcal{D} \phi\, p(\data | \phi) p(\phi | \ell)$. This quantity, commonly called the ``evidence,'' forms the basis for Bayesian model selection \cite{Bialek:1996, Nemenman:2002, MacKay:2003:Occam, Balasubramanian:1997}. 

For the problem in hand, the evidence vanishes when $\alpha > 1$ regardless of the data. The reason for this is that $p(Q | \ell)$ is an improper prior; see Appendix C. However, the evidence ratio $E = p(\data | \ell) / p(\data | \infty)$ is finite for all $\ell > 0$. Using a Laplace approximation, which is valid for large $N$, we find that 
\bea
E(\ell) = e^{S_\infty [\phi_\infty] - S_\ell[\phi_\ell]} \sqrt{ \frac{\det_{\ker} [ e^{-\phi_\infty} ] \det_{\row} [ L^{2 \alpha} \Delta^\alpha] }{\eta^{-\alpha} \det [ L^{2 \alpha} \Delta^\alpha + \eta e^{-\phi_\ell} ] } },~~~~~\label{eq:evidence_ratio}
\eea
where $\eta = N (L/\ell)^{2 \alpha}$. Here the subscripts ``row'' and ``ker'' indicate restriction to the row space and kernel of $\Delta^\alpha$, respectively; the $e^{-\phi_\ell}$ terms inside the determinants stand for matrices that have the values $e^{-\phi_\ell(x)}$ (for all $x$s) arrayed along the main diagonal and zeros everywhere else. See Appendix C for the derivation of \eq{eq:evidence_ratio}.

% Give perturbation theory result
By construction, the evidence ratio $E(\ell)$ approaches unity in the large $\ell$ limit. Whether this limiting value is approached from above or below is relevant to the question of whether $\ell = \infty$ is optimal, and thus whether the MaxEnt hypothesis is consistent with the data. Using perturbation theory about $\eta = 0$ ($\ell = \infty$), we find that 
\bea
\ln E = K \eta + O(\eta^2),
\eea
where the coefficient $K$ is \footnote{See SM for a derivation of \eq{eq:K}.} 
\bea
K &=&   \sum_{\substack{i > \alpha \\ {~}}} \frac{N v_i^2 - z_{ii}}{2 \lambda_i} + \sum_{\substack{i > \alpha \\ j \leq \alpha}} \frac{z_{ij}^2 + v_i z_{ijj}}{2 \lambda_i \zeta_j} -  \sum_{\substack{i > \alpha \\ j,k \leq \alpha}} \frac{v_i z_{ij} z_{jkk} }{2 \lambda_i \zeta_j \zeta_k}.~~~~~~~\label{eq:K}
\eea
Here, $\lambda_i$ and $\psi_i(x)$ ($i = 1, 2, \ldots$) denote the eigenvalues and eigenfunctions of $L^{2 \alpha} \Delta^\alpha$ and are indexed so that $\lambda_i = 0$ for $i \leq \alpha$. The eigenfunctions are normalized so that $\int dx\, L^{-1} \psi_i \psi_j = \delta_{ij}$, and in the degenerate subspace ($i,j \leq \alpha$) they are oriented so that $\int dx\,Q_\infty \psi_i \psi_j = \delta_{ij} \zeta_j$ for some positive real numbers $\zeta_j$. The other indexed quantities are $v_i = \int dx\,(Q_\infty - R) \psi_i$, $z_{ij} = \int dx\,Q_\infty \psi_i \psi_j$, and $z_{ijk} =  \int dx\,Q_\infty  \psi_i \psi_j \psi_k$.

\eq{eq:K} provides a plug-in formula that can be used to  assess the validity of the MaxEnt hypothesis. If $K > 0$, there is guaranteed to be a finite value of $\ell$ that has a larger evidence ratio than $\ell = \infty$. In this case the MaxEnt estimate is guaranteed to be non-optimal. On the other hand, if $K < 0$, then $\ell = \infty$ is a local optimum that may or may not be a global optimum as well. Numerical computation of $E$ over all values of $\ell$ is thus needed to resolve whether the MaxEnt hypothesis provides the best explanation of the data in hand. 

%
% Figures
%

% Figure 1
\begin{figure*}
\includegraphics{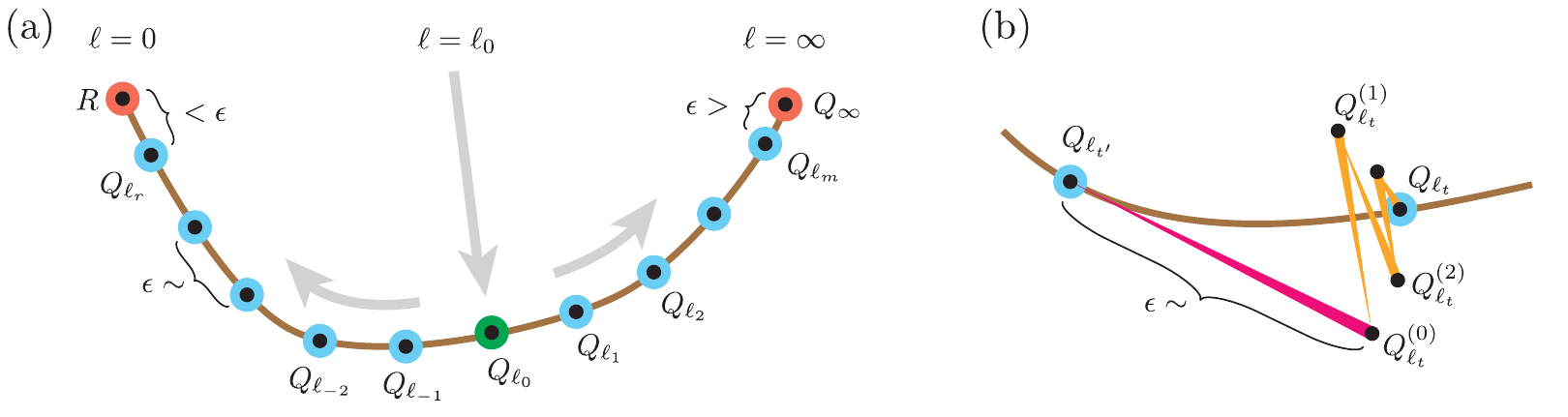}	
\caption{(Color) Illustration of the predictor-corrector homotopy algorithm. (a) The MAP curve (brown) is approximated using finite set of densities $\set{R, Q_{\ell_r}, \cdots, Q_{\ell_{-2}}, Q_{\ell_{-1}}, Q_{\ell_0}, Q_{\ell_1}, Q_{\ell_2}, \ldots, Q_{\ell_m}, Q_{\infty}}$. First the MAP density at an intermediate length scale $\ell_0 = L/\sqrt{G}$ is computed. A predictor-corrector algorithm is then used to extend the set of MAP densities outward to larger and to smaller values of $\ell$. These $\ell$ values are chosen so that neighboring MAP densities lie within a geodesic distance of $\lesssim \epsilon$ of each other. (b) Each step $Q_{\ell_{t'}} \to Q_{\ell_t}$ has two parts. First, a predictor step (magenta) computes a new length scale $\ell_t$ and an approximation $Q_{\ell_t}^{(0)}$ of $Q_{\ell_t}$. A series of corrector steps $Q_{\ell_t}^{(0)} \to Q_{\ell_t}^{(1)} \to Q_{\ell_t}^{(2)} \cdots$ (orange) then converges to $Q_{\ell_t}$. \label{fig:algorithm}}
\end{figure*}

% Figure 2
\begin{figure}[!t]
\includegraphics{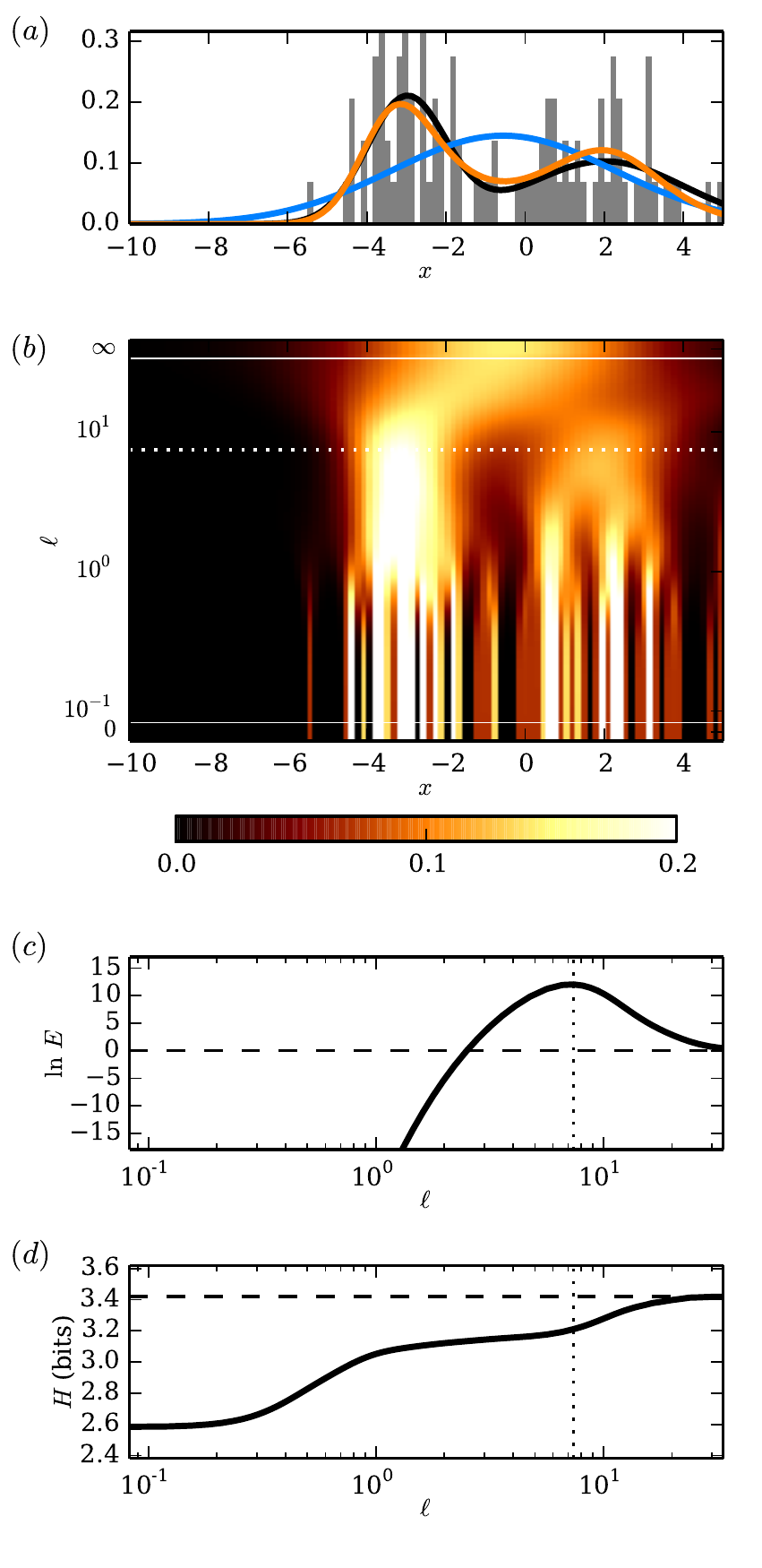}
\caption{(Color) Density estimation without boundary conditions using the $\alpha = 3$ field theory prior. (a) N = 100 data points were drawn from a simulated density $Q_\true$ (black) and binned at $G = 100$ grid points. The resulting histogram (gray) is shown along with the field theory density estimate $Q_{\ell^*}$ (orange) and the corresponding MaxEnt estimate $Q_\infty$ (blue). (b) The heat map shows the densities $Q_\ell$ interpolating between the MaxEnt density at $\ell = \infty$ and the data histogram at $\ell = 0$. (c) The log evidence ratio $E$ is shown as a function of $\ell$. (d) The differential entropy $H = -\int dx\,Q_\ell \ln Q_\ell$ \cite{Cover:2006} is shown as a function of $\ell$; the entropy at $\ell = \infty$ is indicated by the dashed line. Dotted lines in (b-d) mark the optimal length scale $\ell^*$. \label{fig:one_dim}}
\end{figure}

 % Figure 3
 \begin{figure*}[!t]
 \includegraphics{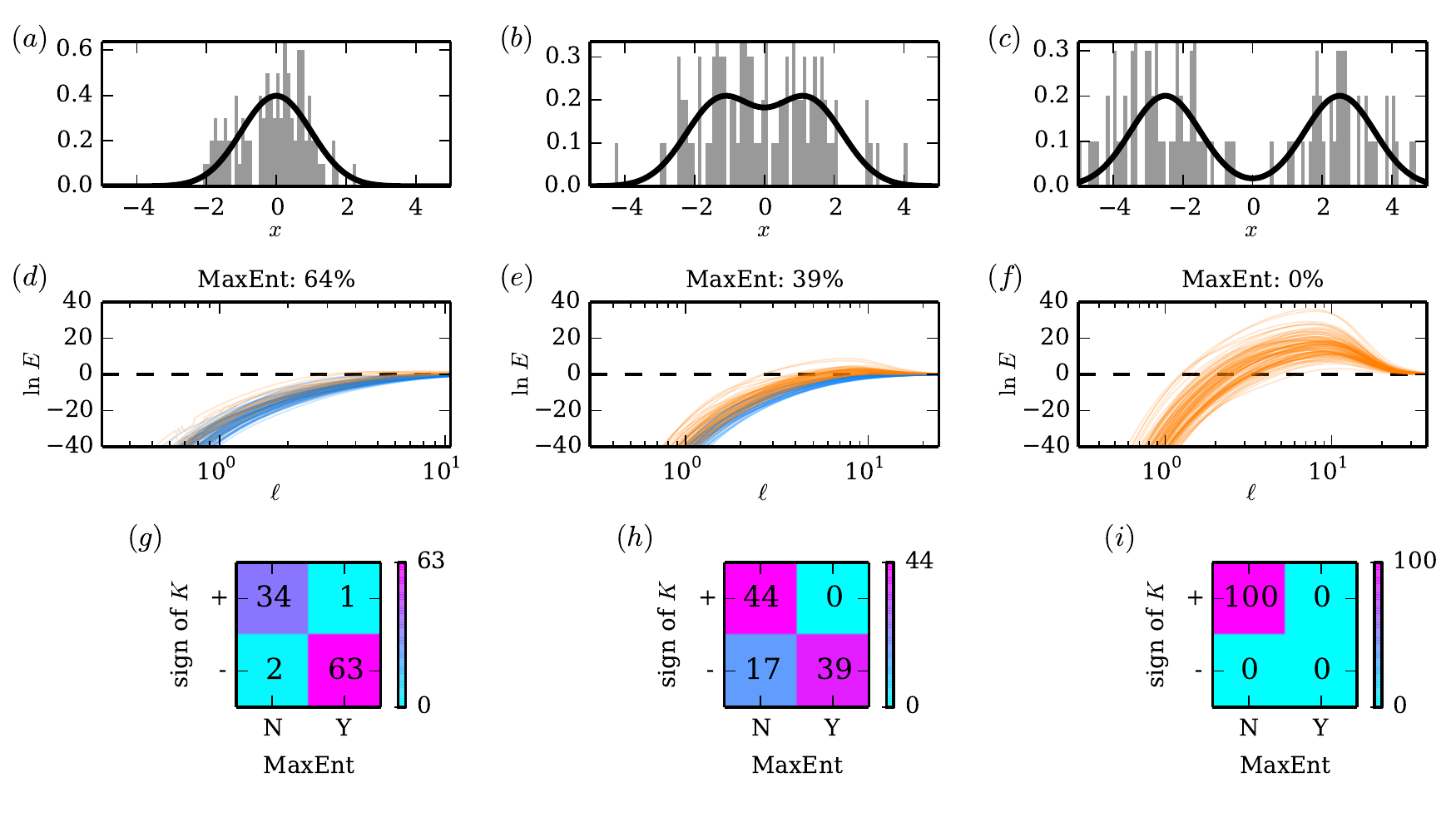}
 \caption{(Color) The optimal estimated density for any particular data set might or might not have maximum entropy. Panels (a-c) show three different choices for $Q_\true$ (black), along with a histogram (gray) of $N = 100$ data points binned at $G = 100$ grid points. In each panel, $Q_\true$ was chosen to be the sum of two equally weighted normal distributions separated by a distance of (a) 0, (b) 2.5, or (c) 5. Panels (d-f) show the evidence ratio curves computed for 100 data sets respectively drawn from the $Q_\true$ densities in (a-c). Blue curves indicate $\ell^* = \infty$; orange curves indicate finite $\ell^*$. Titles in (d-f) give the percentage of data sets for which $\ell^* = \infty$ was found. The heat maps shown in panels (g-i) report the number of simulated data sets for which the $K$ coefficient was positive or negative and for which the MaxEnt density was or was not recovered.}
 \end{figure*}

%
% Describe the discrete representation
%
\section{Discrete space representation}

In this section we retrace the entire analysis above in the discrete representation, i.e., where the continuous $x$-interval is replaced by an evenly spaced set of $G$ grid points. This discrete representation is necessary for the computational implementation of our field theoretic density estimation method.  Happily, our main findings above are seen to hold exactly upon discretization. This discrete representation also sheds light on how the bilateral Laplacian differs from the standard Laplacian and why this difference eliminates the need for boundary conditions. 

We consider $G$ grid points evenly spaced throughout the interval $[x_\mathrm{min}, x_\mathrm{max}]$. Specifically, we place grid points at
\bea
x_i = x_\mathrm{min} + h \left( n - \frac{1}{2} \right),~~~n = 1, 2, \ldots G
\eea
where $h = L/G$ is the grid spacing. In moving to this discrete representation, functions of $x$ become $G$-dimensional vectors with elements denoted by the subscript $n$. For instance, the field $\phi(x)$ becomes a vector with elements $\phi_n$. Integrals become sums, i.e.,
\bea
\int dx\, f(x) \to h \sum_{n=1}^G f_n,
\eea
and path integrals over $\phi$ become $G$-dimensional integrals over the elements $\phi_n$, i.e.,
\bea
\int \mathcal{D} \phi \to \int_{-\infty}^\infty d \phi_1 \int_{-\infty}^\infty d \phi_2 \cdots \int_{-\infty}^\infty d \phi_G.
\eea

% Discuss the discrete representation of derivatives and the resulting dimensionality reduction
We denote differential operators in this discrete representation with a subscript $G$. The derivative operator, $\partial_G$, becomes a $(G-1) $ by $G$ matrix having elements $(\partial_G)_{nm} = h^{-1} (-\delta_{n,m} + \delta_{n+1,m})$. For instance, setting $G = 8$ gives the $7 $ by $ 8$ matrix,
\bea
\partial_8 &=& \frac{1}{h} \mat{rrrrrrrr}{
-1 & 1 & 0 & 0 & 0 & 0 & 0 & 0 \\
0 & -1 & 1 & 0 & 0 & 0 & 0 & 0 \\
0 & 0 & -1 & 1 & 0 & 0 & 0 & 0 \\
0 & 0 & 0 & -1 & 1 & 0 & 0 & 0 \\
0 & 0 & 0 & 0 & -1 & 1 & 0 & 0 \\
0 & 0 & 0 & 0 & 0 & -1 & 1 & 0 \\
0 & 0 & 0 & 0 & 0 & 0 & -1 & ~\,1 \\
}.~~~~~
\eea
Similarly, the standard $\alpha$-order Laplacian becomes a $(G-2 \alpha) $ by $ G$ matrix, given by $(-1)^\alpha \partial_{G-2\alpha+1} \cdots \partial_{G-1} \partial_G$. For example, choosing $\alpha = 3$ and $G = 8$ yields the a $2 $ by $ 8$  Laplacian matrix
\bea
-\partial^6_8 &=& \frac{1}{h^6} \mat{rrrrrrrr}{
-1 & 6 & -15 & 20 & -15 & 6 & -1 & 0 \\
0 & -1 & 6 & -15 & 20 & -15 & 6 & -1
}.~~~
\eea
Because $2 \alpha$ elements are eliminated from the vector $\phi_\ell$ upon application of the standard Laplacian, the discrete version of \eq{eq:classic_eom}  provides only $G- 2 \alpha$ equations for the $G$ unknown values of $\phi_\ell$. $2 \alpha$ additional constraints, typically provided in the form of boundary conditions, are thus needed to obtain a unique solution. 

By contrast, the $\alpha$-order bilateral Laplacian is represented by the $G $ by $ G$ matrix $\Delta^\alpha_G = \left( \partial^\alpha_G \right)^\top \partial^\alpha_G$, where $\partial^\alpha_G = \partial_{G-\alpha+1} \cdots \partial_{G-1} \partial_G$. Indeed, again choosing $\alpha = 3$ and $G = 8$ we recover an $8 $ by $ 8$ bilateral Laplacian matrix
\bea
\Delta^3_8 &=& \frac{1}{h^6} \mat{rrrrrrrr}{
1 & -3 & 3 & -1 & 0 & 0 & 0 & 0 \\
-3 & 10 & -12 & 6 & 1 & 0 & 0 & 0 \\
3 & -12 & 19 & -15 & 6 & 1 & 0 & 0 \\
-1 & 6 & -15 & 20 & -15 & 6 & -1 & 0 \\
0 & -1 & 6 & -15 & 20 & -15 & 6 & -1\\
0 & 0 & -1 & 6 & -15 & 19 & -12 & 3 \\
0 & 0 & 0 & -1 & 6 & -12 & 10 & -3 \\
0 & 0 & 0 & 0 & -1 & 3 & -3 & 1 \\
}.~~~~~
\eea
The middle two rows of $\Delta^3_8$ match those of $- \partial^6_8$, reflecting the equivalence of bilateral Laplacians and standard Laplacians in the interior of the $x$-interval. However, $\Delta^3_8$ contains six additional rows, three at either end. These are sufficient to specify a unique solution for the 8 elements of the $\phi_\ell$ vector. More generally, the discrete version of \eq{eq:eom} provides $G$ equations for the $G$ unknown elements of $\phi_\ell$ and is therefore able to specify a unique solution without the imposition of any boundary conditions. 

Using the bilateral Laplacian, we readily define a discretized version of the prior by adopting
\bea
S^0_\ell[\phi] = \frac{\ell^{2 \alpha}}{2 G}  \sum_{n,m} \Delta^\alpha_{nm} \phi_n \phi_m.
\eea
This leads to the posterior action
\bea
S_\ell[\phi] &=& \sum_{n,m} \left\{ \frac{\ell^{2 \alpha}}{2 G} \Delta^\alpha_{nm} \phi_n \phi_m  \right. \\
& & \left. ~~~~~~~+ \frac{N L}{G} R_n \phi_n \delta_{nm} + \frac{N}{G} e^{-\phi_n} \delta_{nm} \right\}, 
\eea
where $R_n$ is value of the data histogram at grid point $n$, i.e., the fraction of data points discretized to grid point $n$, divided by bin width $h$. 

The corresponding equation of motion is
\bea
\ell^{2 \alpha} \sum_m \Delta^\alpha_{nm} \phi_{\ell m} + N L R_n - N e^{-\phi_{\ell n}} = 0. \label{eq:eom_grid}
\eea
The kernel of $\Delta^\alpha_G$ is spanned by vectors $\phi$ having the polynomial form $\phi_n = \sum_{b = 0}^{\alpha-1} a_b x_n^b$. The analogous moment-matching behavior therefore holds exactly in the discrete representation, i.e.,
\bea
h \sum_{n = 1}^G Q_{\ell n}\,x_n^k = h \sum_{n = 1}^G R_n\,x_n^k \label{eq:discrete_moment_matching}
\eea
where $Q_\ell$ is related to $\phi_\ell$ via \eq{eq:MAP_field_to_MAP_density}. In the $\ell \to \infty$ limit, the MAP density $Q_\infty$ again has the analogous form 
\bea
Q_{\infty n} &=& \frac{1}{L} \exp \left( - \sum_{k=0}^{\alpha-1} a_k x_n^k \right)
\eea
where the coefficients $a_k$ are chosen to satisfy \eq{eq:discrete_moment_matching}. Thus, the connection to the MaxEnt density estimate remains intact upon discretization. 

The derivation of the evidence ratio in \eq{eq:evidence_ratio} follows through without modification. The only change is that the functional determinants now become determinants of finite-dimensional matrices. The derivation of the $K$ coefficient in \eq{eq:K} also follows in a similar manner; the only change to \eq{eq:K} is that the the index $i$ now ranges from 1 to $G$, not 1 to $\infty$. 

%
% Describe the computations
%
\section{Computing density estimates}

To compute density estimates using this field theory approach, we work in the discrete representation described in the previous section. First the user specifies the number of grid points $G$ as well as a bounding box $[x_\mathrm{min}, x_\mathrm{max}]$ for the data. MAP densities $Q_\ell$ are then computed at a finite set of length scales $\set{0, \ell_r, \cdots, \ell_{-2}, \ell_{-1}, \ell_0, \ell_1, \ell_2, \cdots, \ell_m, \infty}$, as illustrated in Fig.\ 1a. This ``string of beads'' approximation to the MAP curve allows us to evaluate the evidence ratio $E$ over all length scales and, to finite precision, identify the  length scale $\ell^*$ that maximizes $E$. 

% Describe the homotopy algorithm
This approximation of the MAP curve is computed using a predictor-corrector homotopy algorithm \cite{Allgower:1990}. An overview of this algorithm is now given. Please refer to Appendix E for algorithm details. I note that this algorithm provides more transparent precision bounds on the computed $Q_\ell$ densities than does the previously reported algorithm of \cite{Kinney:2014}.  

% Give overview of the algorithm
First, an intermediate length scale $\ell_0$ is chosen and the corresponding MAP density $Q_{\ell_0}$ is computed. This density, $Q_{\ell_0}$, serves as the starting point from which to trace MAP curve towards both larger and  smaller length scales (Fig.\ 1a). The algorithm then proceeds in both directions, stepping from length scale to length scale and updating the MAP density at each step. 

% Describe tracing of the MAP curve
During each step, the subsequent length scale is chosen so that the corresponding MAP density is sufficiently similar to the MAP density at the preceding length scale.  Specifically, in stepping from $\ell_{t'}$ to $\ell_t$, the algorithm chooses $\ell_t$ so that the geodesic distance $D_\mathrm{geo}$ (see \cite{Skilling:2007,Kinney:2014}) between $Q_{\ell_{t'}}$ and $Q_{\ell_t}$ matches a user-specified tolerance $\epsilon$, i.e.,
\bea
D_{\rm geo}[Q_{\ell_t}, Q_{\ell_{t'}}] \equiv 2 \cos^{-1} \left( \int dx\, \sqrt{Q_{\ell_t} Q_{\ell_{t'}}} \right) \, \lesssim \, \epsilon. \label{eq:distance_criterion}
\eea
The value $\epsilon = 10^{-2}$ was used for the computations described below and shown in Figs.\ 2 and 3. Stepping in the decreasing $\ell$ direction is terminated at a length scale $\ell_r$ such that $D_\mathrm{geo}[Q_{\ell_r}, R] < \epsilon$. Similarly, stepping in the increasing $\ell$ direction is terminated at a length scale $\ell_m$ such that $D_\mathrm{geo}[Q_{\ell_m}, Q_\infty] < \epsilon$; the MaxEnt density $Q_\infty$ is computed at the start of the algorithm essentially as described by Ormoneit and White \cite{Ormoneit:1999}.

% Describe the predictor and corrector steps
Each step along the MAP curve is accomplished in two parts (Fig.\ 1b). Given the MAP density $Q_{\ell_{t'}}$ at length scale $\ell_{t'}$, a ``predictor step'' is used to compute both the next length scale $\ell_t$ as well as an approximation $Q_{\ell_t}^{(0)}$ to the corresponding MAP density $Q_{\ell_t}$. The repeated application of a ``corrector step'' is then used to compute a series of densities $Q_{\ell_t}^{(1)}, Q_{\ell_t}^{(2)}, \dots$ that converges to $Q_{\ell_t}$.

% Describe why this algorithm is expected to work well
If the numerics are properly implemented, this predictor-corrector algorithm is guaranteed to identify the correct MAP density $Q_\ell$ at each of the chosen length scales $\ell$. This is because the action $S_\ell[\phi]$ is strictly convex in $\phi$ and therefore has a unique minimum (as was shown in section III). The distance criteria in \eq{eq:distance_criterion} further ensures that the stepping procedure does not drastically overstep $\ell^*$. It is also worth noting that, because $\Delta_G^\alpha$ is sparse, these predictor and corrector steps can be sped up by using numerical sparse matrix methods. 

%
% Example analyses
%
\section{Example analyses}

Fig.\ 2 provides an illustrated example of this density estimation procedure in action. Starting from a set of sampled data (Fig.\ 2a, gray), the homotopy algorithm computes the MAP density $Q_\ell$ at a set of length scales spanning the interval $\ell \in [0, \infty]$ (Fig.\ 2b). The evidence ratio $E$ is then computed at each of these chosen length scales using \eq{eq:evidence_ratio}, and the length scale $\ell^*$ that maximizes $E$ is identified (Fig.\ 2c). $Q_{\ell^*}$ is then reported as the best estimate of the underlying density (Fig.\ 2a, orange). If one further wishes to report ``error bars'' on this estimate, other plausible densities $Q$ can be drawn from the posterior $p(Q | \data)$ as described in \cite{Kinney:2014}. 

The optimal length scale $\ell^*$ may or may not be infinite. If $\ell^* = \infty$, then $Q_{\ell^*}$ is the MaxEnt estimate that matches the first $\alpha - 1$ moments of the data. On the other hand, if $\ell^*$ is finite as in Fig.\ 2, then $Q_{\ell^*}$ will have lower entropy than the MaxEnt estimate (Fig.\ 2d) while still exactly matching the first $\alpha-1$ moments of the data. This  reduced entropy reflects the use of addition information in the data beyond the first $\alpha-1$ moments. It should be noted that $\ell^*$ is never zero due to a vanishing Occam factor in this limit. 

The density estimation procedure proposed in this paper thus provides an automatic test of the MaxEnt hypothesis. It can therefore be used to test whether $Q_\true$ has a hypothesized functional form.  For example, using $\alpha = 3$ we can test whether our data was drawn from a Gaussian distribution. This is demonstrated in Fig.\ 3. In these tests, when data was indeed drawn from a Gaussian density, $\ell^* = \infty$ was obtained about $64\%$ of the time (Fig.\ 3a and 3d). By contrast, when data was drawn from a mixture of two Gaussians, the fraction of data sets yielding $\ell^* = \infty$ decreased sharply as the separation between the two Gaussians was increased (Figs.\ 3b, 3c, 3e, and 3f). In a similar manner, this density estimation approach can be used to test other functional forms for $Q_\true$, either by using the bilateral Laplacian of different order $\alpha$, or by replacing the bilateral Laplacian with a differential operator having a kernel spanned by other functions whose expectation values one wishes to match to the data. 

The method used to select $\ell^*$ both here and in previous work \cite{Bialek:1996, Nemenman:2002} is sometimes referred to as ``empirical Bayes'': for $\ell^*$ we choose the value of $\ell$ that maximizes $p(\data | \ell)$. By contrast, \cite{Kinney:2014} used a fully Bayesian approach by positing a Jeffreys prior $p(\ell)$ then choosing the length scale $\ell$ that maximizes $p(\data, \ell) \sim p(\data | \ell) p(\ell)$. It can be reasonably argued that the empirical Bayes method adopted here is less sensible than the fully Bayesian approach. However, in the fully Bayesian approach the assumed prior $p(\ell)$ obscures the large $\ell$ behavior of the evidence ratio $E$. This large $\ell$ behavior is nontrivial and potentially useful. 

As shown in section V, the behavior of $E$ in the large $\ell$ limit is governed by the $K$ coefficient defined in \eq{eq:K}. The sign of the $K$ coefficient can therefore be used to assess the MaxEnt hypothesis without having to compute $E$ at every length scale. This suggestion is supported by the simulations shown in Fig.\ 3. Here, the sign of $K$ (positive or negative) performed well as a proxy for whether the MaxEnt estimate was recovered (no or yes, respectively) from a full computation of the MAP curve; see Figs.\ 3g \footnote{The one trial reported in Fig.\ 3g for which $\ell^* = \infty$ even though $K > 0$ is due to difficulties with the numerics in the $\ell \to \infty$ limit.}, 3h, and 3i. These results suggest that the $K$ coefficient, for which \eq{eq:K} provides an analytic expression, might allow an expedient test of the MaxEnt hypothesis when computation of the entire MAP curve is less feasible, e.g., in higher dimensions. 

%
% Summary and Discussion
%
\section{Summary and discussion} 

% Review technical details of prior literature
Bialek et al.\ \cite{Bialek:1996} showed that probability density estimation can be formulated as a Bayesian inference problem using field theory priors.  Among other results, \cite{Bialek:1996} derived the action in \eq{eq:action} and showed how to use a Laplace approximation of the evidence to select the optimal smoothness length scale \footnote{See Appendix F for a discussion of earlier related work on the ``maximum penalized likelihood'' formulation of the density estimation problem and how it relates to the Bayesian field theory approach.}. However, periodic boundary conditions were imposed on candidate densities in order to transform the standard Laplacian into a Hermitian operator. The MaxEnt density estimate typically violates these boundary conditions, and as a result the ability of Bayesian field theory to subsume MaxEnt density estimation went unrecognized in \cite{Bialek:1996} and in follow-up studies \cite{Nemenman:2002,Kinney:2014}. 

% Summarize results presented here
Here we have seen that boundary conditions on candidate densities are unnecessary. The bilateral Laplacian, defined in \eq{eq:bl_def}, is a Hermitian operator that imposes no boundary conditions on functions in its domain, yet is equivalent to the standard Laplacian in the interior of the interval of interest. Using the bilateral Laplacian of various orders to define field theory priors, we recovered standard MaxEnt density estimates in cases where the smoothness length scale was infinite. We also obtained a novel criterion for judging the appropriateness of the MaxEnt hypothesis on individual data sets. 

% This applies to any MaxEnt density estimation problem. 
More generally, Bayesian field theories can be constructed for any set of moment-matching constraints. One need only replace the bilateral Laplacian in the above equations with a differential operator that has a kernel spanned by the functions whose mean values one wishes to match to the data. The resulting field theory will subsume the corresponding MaxEnt hypothesis, and thereby allow one to judge the validity of that hypothesis. Analogous approaches for estimating discrete probability distributions can be formulated by replacing integrals over $x$ with sums over states. 

% Many problems have now been overcome
The elimination of boundary conditions removes a considerable point of concern with using Bayesian field theory for estimating probability densities. As demonstrated here and in \cite{Kinney:2014}, the necessary computations are readily carried out in one dimension. One issue not investigated here -- the large $N$ assumption used to compute the evidence ratio -- can likely be addressed by using Feynman diagrams in the manner of \cite{Ensslin:2009}.

% Discuss issue of choosing the prior
In the author's opinion, the problem of how to choose an appropriate prior appears to be the primary issue standing in the way of a definitive practical solution to the density estimation problem in 1D. It is not hard to imagine different situations that would call for qualitatively different priors, but understanding which situations call for which priors will require further study. The author is optimistic, however, that the variety of priors needed to address most of the 1D density estimation problems typically encountered might not be that large.

% Discuss density estimation in higher dimensions
This field theory approach to density estimation readily generalizes to higher dimensions -- at least in principle. Additional care is required in order to construct field theories that do not produce ultraviolet divergences \cite{Bialek:1996}, and the best way to do this remains unclear. The need for a very large number of grid points also presents a substantial practical challenge. Grid-free methods, such as those used by \cite{GoodGaskins:1971, Holy:1997}, may provide a way forward. 

\begin{acknowledgments}
I thank Gurinder Atwal, Curtis Callan, William Bialek, and Vijay Kumar for helpful discussions. Support for this work was provided by the Simons Center for Quantitative Biology at Cold Spring Harbor Laboratory.
\end{acknowledgments}

%
% Appendix: Derivation of the action
%
\appendix
\section{Derivation of the action}

Our derivation of the action $S_\ell[\phi]$ in \eq{eq:action} of the main text is essentially that used in \cite{Kinney:2014}. This derivation is not entirely straight-forward, however, and the details of it have yet to be reported. 

The prior $p(\phi | \ell)$, which is defined by the action $S^0_\ell$ in \eq{eq:prior}, is improper due to the differential operator $\Delta^\alpha$ having an $\alpha$-dimensional kernel; see section III. To avoid unnecessary mathematical difficulties, we can render $p(\phi | \ell)$ proper by  considering a regularized form of the action
\bea
S_\ell^0[\phi] &=& \int \frac{dx}{L} \frac{1}{2} \phi \left[ \ell^{2 \alpha} \Delta^\alpha + \epsilon \right] \phi, \label{eq:regularized_action}
\eea
where $\epsilon$ is an infinitesimal positive number. All quantities of interest, of course, should be evaluated in the $\epsilon \to 0$ limit. 

The corresponding prior over $Q$ is
\bea
p(Q | \ell) = \int_{-\infty}^{\infty} d\phi_c\, p(\phi | \ell)  =  \sqrt{\frac{2 \pi}{\epsilon}} \frac{e^{-S^0_\ell[\phi_{nc}]}}{Z_\ell^0}.
\eea
Here we have decomposed the field $\phi$ using
\bea
\phi(x) = \phi_{nc}(x) + \phi_c
\eea
where $\phi_c$ is the constant Fourier component of $\phi$ and $\phi_{nc}(x)$ is the non-constant component of $\phi$. The likelihood of $Q$ given the data is given by
\bea
p(\data | Q) = \prod_{n=1}^N Q(x_n). %= (Z_\ell^0)^{-N}\, e^{ - \sum_i S_\ell^0 [\phi(x_i)] }
\eea
Using the identity
\bea
a^{-N} = \frac{N^N}{\Gamma(N)} \int_{-\infty}^\infty du\, e^{-N \left( u + a e^{-u} \right)},
\eea
which holds for any positive numbers $a$ and $N$, we find that the likelihood of $Q$ can be expressed as
\bea
p(\data | Q) 
%&=& \exp \left( - \sum_i \phi_{nc}(x_i) \right)  \frac{1}{L^{N}} \left[ \int \frac{dx}{L}\, e^{-\phi_{nc}(x)} \right]^{-N} \\
%&=& \exp \left( - \int dx\,N R \phi_{nc} \right)  \frac{N^N}{L^{N} \Gamma(N)} \int_{-\infty}^\infty d \phi_c \exp \left( - N \left[ \phi_c + \int \frac{dx}{L}\, e^{-\phi_{nc}(x) - \phi_c} \right] \right) ~~~~~\\
%= \frac{N^N}{L^{N} \Gamma(N)} \int_{-\infty}^\infty d \phi_c \exp \left( - \int \frac{dx}{L} \left\{ N L R \phi + N e^{-\phi} \right\} \right).
= \frac{N^N}{L^{N} \Gamma(N)} \int_{-\infty}^\infty d \phi_c e^{ - \int \frac{dx}{L} \left\{ N L R \phi + N e^{-\phi} \right\} }~~~.
\eea
The prior probability of $Q$ and the data together is therefore given by
\bea
p(\data, Q | \ell) %&=& p(\data | Q)\, p(Q | \ell) \\
%&=& \frac{N^N}{L^{N} \Gamma(N)} \sqrt{\frac{2 \pi}{\epsilon}}  \int_{-\infty}^\infty d \phi_c \exp \left( - \int \frac{dx}{L} \left\{ N L R \phi - N e^{-\phi} \right\} \right)  \frac{1}{Z_\ell^0} \exp \left( - \int \frac{dx}{L} \frac{\ell^{2 \alpha}}{2} (\partial^\alpha \phi_{nc})^2 \right)\\
%&=& \frac{N^N}{L^{N} \Gamma(N)} \sqrt{\frac{2 \pi}{\epsilon}}  \frac{1}{Z_\ell^0} \int_{-\infty}^\infty d \phi_c\, \exp \left( - \int \frac{dx}{L} \left\{ \frac{\ell^{2 \alpha}}{2} (\partial^\alpha \phi)^2  + N L R \phi + N e^{-\phi} \right\} \right) ~~~~~~\\
&=& \frac{N^N}{L^{N} \Gamma(N)} \sqrt{\frac{2 \pi}{\epsilon}}  \frac{1}{Z_\ell^0} \int_{-\infty}^\infty d \phi_c\, e^{-S_\ell[\phi]},~~~~
\eea
where $S_\ell[\phi]$ is the action from \eq{eq:action}.  
%\bea
%S_\ell[\phi] = \int \frac{dx}{L} \left\{ \frac{\ell^{2 \alpha}}{2} (\partial^\alpha \phi)^2  + N L R \phi + N e^{-\phi} \right\}.
%\eea

The ``evidence'' for $\ell$ -- i.e., the probability of the data given $\ell$ -- is therefore given by,
\bea
p(\data | \ell) %&=& \int \mathcal{D}Q\, p(\data, Q | \ell) \\
%&=& \frac{N^N}{L^{N} \Gamma(N)} \sqrt{\frac{2 \pi}{\epsilon}}  \frac{1}{Z_\ell^0}  \int \mathcal{D} \phi\, e^{-S_\ell[\phi]} \\
&=& \frac{N^N}{L^{N} \Gamma(N)} \sqrt{\frac{2 \pi}{\epsilon}} \frac{Z_\ell}{Z_\ell^0}, \label{eq:evidence_formula}
\eea
where $Z_\ell$ is the partition function from \eq{eq:posterior_partition_function}. The posterior distribution over $Q$ is then given by Bayes's rule:
\bea
p(Q | \data, \ell) &=& \frac{p(\data, Q | \ell)}{p(\data| \ell)} \\
&=& \int_{-\infty}^{\infty} d\phi_c\, \frac{e^{-S_\ell[\phi]}}{Z_\ell}. \label{eq:not_a_problem}
\eea
This motivates us to \emph{define} the posterior distribution over $\phi$ as
\bea
p(\phi | \data, \ell) &\equiv& \frac{e^{-S_\ell[\phi]}}{Z_\ell}. \label{eq:posterior_def}
\eea
This definition of $p(\phi | \data, \ell)$ is consistent with the formula for $p(Q | \data, \ell)$ obtained in \eq{eq:not_a_problem}, in that
\bea
p(Q | \data, \ell) &=& \int_{-\infty}^\infty d \phi_c\, p(\phi | \data, \ell). 
\eea
However, \eq{eq:posterior_def} violates Bayes's rule, since
\bea
p(\phi | \data, \ell) \neq \frac{p(\data, \phi | \ell)}{p(\data | \ell)}.
\eea 
This is not a problem, however, since $\phi$ itself is not directly observable; only $Q$ is observable. Put another way, \eq{eq:posterior_def} violates Bayes's rule only in the way that it specifies the behavior of $\phi_c$. This constant component of $\phi$, however, has no effect on $Q$. 

Note that all of the above calculations have been exact; no large $N$ approximation was used. This contrasts with prior work \cite{Bialek:1996, Nemenman:2002}, which used a large $N$ Laplace approximation to derive the formula for $S_\ell[\phi]$. Also note that the regularization parameter $\epsilon$ has vanished in the formulas for the posterior distributions $p(Q | \data, \ell)$ and $p(\phi | \data, \ell)$. However, this parameter still appears in the formula for the evidence $p(\data | \ell)$, both explicitly and implicitly through the value of $Z_\ell^0$. 

%
% Appendix: Spectrum of the bilateral Laplacian
%null space

\section{Spectrum of the bilateral Laplacian}

\begin{figure}[!t]
\includegraphics{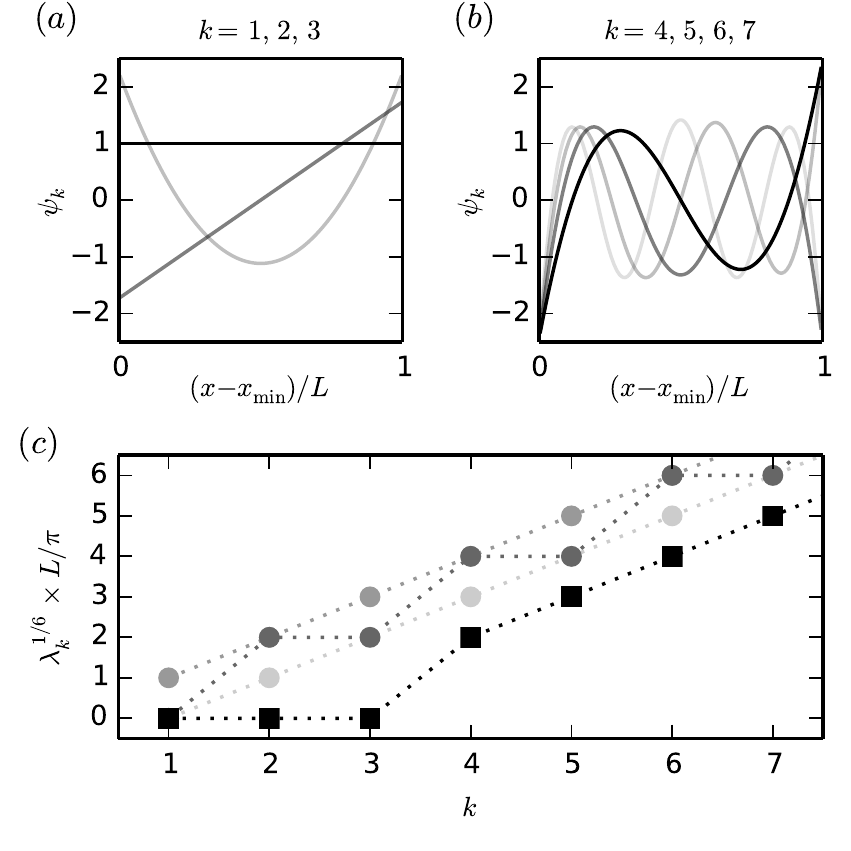}
\caption{Spectrum of the bilateral Laplacian of order $\alpha = 3$. (a) The first three Legendre polynomials provide an orthonormal basis for the kernel of $\Delta^3$. These choices for $\psi_1$, $\psi_2$, and $\psi_3$ are plotted with decreasing opacity. (b) All other eigenfunctions are nontrivial linear combinations of factors of the form $\exp(i \kappa x)$. This behavior is illustrated by the basis functions $\psi_4$, $\psi_5$, $\psi_6$, and $\psi_7$, which are shown in decreasing opacity. (c) The rank-ordered eigenvalues of $\Delta^3$ lie at or below those of the standard Laplacian with any choice of boundary conditions. Shown are the eigenvalues of $\Delta^3$ (black squares), together with the eigenvalues of $- \partial^6$ with periodic, Dirichlet, or Neumann boundary conditions (dark, medium, and light gray circles respectively). \label{fig:spectrum}}
\end{figure}

In the continuum limit, $\Delta^\alpha = (\partial^\alpha)^\top \partial^\alpha$ remains manifestly Hermitian and therefore possesses a complete orthonormal basis of eigenfunctions. We now consider the spectrum of this operator. In what follows we will make use the ket notion of quantum mechanics, primarily as a notational convenience. For any two functions $f$ and $g$ and any Hermitian operator $H$, we define 
\bea
\braket{f | H | g} = \int \frac{dx}{L} f^* H g.
\eea
Note the convention of dividing by $L$ in the inner product integral. This allows us to take inner products without altering units. 

From
\bea
\braket{\phi | \Delta^\alpha | \phi} = \int \frac{dx}{L} ~|\partial^\alpha \phi|^2 \ge 0, \label{eq:possemidef}
\eea
we see that $\Delta^\alpha$ is positive semi-definite. Equality in \eq{eq:possemidef} obtains if and only if $\phi$ is a polynomial of order $\alpha-1$; such polynomials therefore comprise the null space of $\Delta^\alpha$. 

More generally, any solution to the eigenvalue equation $\Delta^\alpha \psi = \lambda \psi$ implies that $\lambda \braket{\phi | \psi} = \braket{\phi | \Delta^\alpha | \psi}$ for any test function $\phi$. Integrating this by parts gives
\bea
\lambda \int_{x_\mathrm{min}}^{x_\mathrm{max}} dx~\phi^* \psi &=& \sum_{b=0}^{\alpha-1} \left[ (-1)^b (\partial^{\alpha -b-1} \phi^*) (\partial^{\alpha+b} \psi) \right]_{x_\mathrm{min}}^{x_\mathrm{max}} ~~~~~~~\\ 
& & + (-1)^\alpha \int_{x_\mathrm{min}}^{x_\mathrm{max}} dx~\phi^* \partial^{2 \alpha} \psi. \label{eq:a}
\eea
Considering test functions $\phi(x)$ who's first $\alpha-1$ derivatives all vanish at the boundary, we see that in the bulk of the interval, $x_\mathrm{min} < x < x_\mathrm{max}$,
\bea
\lambda \psi = (-1)^\alpha \partial^{2 \alpha} \psi.
\eea
Any function of $\Delta^\alpha$ must therefore be an eigenfunction of the standard $\alpha$-order Laplacian, $(-1)^\alpha \partial^{2\alpha}$, as well. Moreover, all boundary terms in \eq{eq:a} must vanish because the values of $\phi$ and its derivatives at the boundary are independent of its integral with any function in the interior. The eigenfunction $\psi$ must therefore have the boundary behavior
\bea
0 = \partial^{\alpha+b} \psi |_{x_\mathrm{min}} = \partial^{\alpha+b} \psi |_{x_\mathrm{max}}~~~{\rm for}~~~0 \leq b < \alpha. ~~~~\label{eq:spectrum_boundary_behavior}
\eea
Note in particular that this behavior is exhibited by polynomials of order $\alpha-1$, which comprise the kernel of $\Delta^\alpha$. On the other hand, if $\lambda > 0$,  the corresponding eigenfunction $\psi$ will consist of $2 \alpha$ terms of the form $\exp[ i \kappa x ]$, where $\kappa = \lambda^{1/2\alpha} e^{i \pi b / \alpha }$ for $b = 0, 1, \dots, 2 \alpha-1$. The coefficients of these terms will be such that the boundary behavior in \eq{eq:spectrum_boundary_behavior} is satisfied. Typically such eigenfunctions will not be purely sinusoidal or purely exponential, but rather will exhibit a nontrivial combination of sinusoidal and exponential behavior (see Fig.\ 4b).

It should be emphasized that the boundary behavior of the eigenfunctions (\eq{eq:spectrum_boundary_behavior}) is not a boundary condition that all functions $\phi$ in the domain of $\Delta^\alpha$ must satisfy. Specifically, although any well-behaved function $\phi$ can be expressed in the eigenbasis via
\bea
\phi = \sum_{k = 0}^\infty c_k \psi_k \label{eq:function_decomposition}
\eea
for some set of coefficients $c_k$, one will typically find that
\bea
\partial^b \phi  \neq \sum_{k = 0}^\infty c_k \partial^b \psi_k \label{eq:derivative_decomposition}
\eea
because the sum on the right hand side of \eq{eq:derivative_decomposition} will not be well-defined. The reason for this is that the convergence criterion  for \eq{eq:function_decomposition} is weaker than that of \eq{eq:derivative_decomposition}, due to the fact that $\psi_k \sim 1$ whereas $\partial^b \psi_k \sim k^b$. Therefore, even though the right hand side of \eq{eq:function_decomposition} will converge for any particular $\phi$, the right and side of \eq{eq:derivative_decomposition} typically will not.

The ordered eigenvalues of the bilateral Laplacian provide a lower bound for the eigenvalues of the standard Laplacian with any set of boundary conditions. To see this, note that we can define  a positive semi-definite Hermitian operator $H_{bc}$ whose kernel is spanned by all functions satisfying a set of specified boundary conditions. Let us denote the standard Laplacian with these boundary conditions as $\Delta_{bc}^\alpha$. Then we can express $\Delta_{bc}$ in terms of the bilateral Laplacian as
\bea
\Delta^\alpha_{bc} = \lim_{A \to \infty} \Delta^\alpha_A
\eea
where
\bea
\Delta^\alpha_A = \Delta^\alpha + A H_{bc}.
\eea
In the $A \to \infty$ limit, a prior defined using $\Delta^\alpha_A$ in place of $\Delta^\alpha$ will restrict candidate fields $\phi$ to those that satisfy the boundary condition specified by $H_{bc}$. 

From first-order perturbation theory, we know that the $k$'th eigenvalue of $\Delta^\alpha_{A+dA}$ is related to that of $\Delta^\alpha_A$ by
\bea
\lambda^{A+dA}_k = \lambda^A_k + dA \braket{\psi^A | H_{bc} | \psi^A}.
\eea
Therefore, the $k$'th eigenvalue of $\Delta_{bc}$ is given by
\bea
\lambda^{bc}_k = \lambda_k + \int_0^\infty dA \braket{\psi^A_k | H_{bc} | \psi^A_k}
\eea
where $\psi^A_k$ is the (appropriately defined) $k$'th eigenvector of the operator $\Delta^\alpha_A$. Because $H_{bc}$ is positive semi-definite, the integral on the right hand side is greater or equal to zero. We therefore see that $\lambda^{bc}_k \geq \lambda_k$ for all $k$, regardless of what the actual boundary conditions are. 

This point is illustrated in Fig.\ 4c, which plots the ordered eigenvalues for the $\alpha=3$ bilateral Laplacian together with the ordered eigenvalues of the standard $\alpha$-order Laplacian with three different types of boundary conditions: periodic boundary conditions, 
\bea
\partial^b \psi |_{x_\mathrm{min}} = \partial^b \psi |_{x_\mathrm{max}},~~~b = 0,1,\ldots, 2\alpha-1;~~~~
\eea
Dirichlet boundary conditions, 
\bea
\partial^{2b} \psi|_{x_\mathrm{min}} = \partial^{2b} \psi|_{x_\mathrm{max}} = 0,~~~b = 0, 1, \ldots, \alpha-1;~~~~~~~~
\eea
and Neumann boundary conditions, 
\bea
\partial^{2b+1} \psi|_{x_\mathrm{min}} = \partial^{2b+1} \psi|_{x_\mathrm{max}} = 0,~~~b = 0, 1, \ldots, \alpha-1.~~~~~~~~
\eea

%
% Appendix: Derivation of the evidence ratio
%
\section{Derivation of the evidence ratio}

We now turn to the task of evaluating the partition functions $Z_\ell$ and $Z_\ell^0$, so that we can compute the evidence $p(\data | \ell)$ using \eq{eq:evidence_formula}. Defining $\Lambda = L^{2 \alpha} \Delta^\alpha$ and $\eta = N(L/\ell)^{2 \alpha}$ and working in the grid representation, we find a Hessian of the form
\bea
\left. \frac{\partial^2 S}{\partial \phi_m \partial \phi_n} \right|_{\phi_\ell} = \frac{\ell^{2 \alpha}}{G L^{2 \alpha}} \left( \Lambda_{mn} + \delta_{mn} e^{-\phi_{\ell n}} \right)
\eea
The corresponding Laplace approximation to the path integral is therefore given by
\bea
Z_\ell &\approx& e^{-S_\ell[\phi_\ell]}  \left\{ \left(\frac{\ell^{2 \alpha}}{2 \pi G L^{2 \alpha}} \right)^G \det  \left[ \Lambda + \eta  e^{- \phi_\ell} \right] \right\}^{-1/2}.
\eea
Note that the operator $\Lambda$ is unitless and has well-defined eigenvalues in the $G \to \infty$ limit. Also note that $\eta$ is unitless. For these reasons, $\eta$ will emerge as a natural perturbation parameter in the $\ell \to \infty$ limit. 

Evaluating the partition function $Z_\ell^0$ requires more care because the regularized form of the action $S_\ell^0$, given in \eq{eq:regularized_action}, has to be used in order for the equations we derive to make sense. We find that
\bea
Z_\ell^0 &=&  \left\{  \left(\frac{\ell^{2 \alpha}}{2 \pi G L^{2 \alpha}} \right)^G \det \left[ \Lambda + \frac{\eta \epsilon}{N} \right] \right\}^{-1/2} \label{eq:before_degen_pert} \\
&=& \left\{  \left(\frac{\ell^{2 \alpha}}{2 \pi G L^{2 \alpha}} \right)^G N^{-\alpha} \eta^{\alpha} \epsilon^{\alpha} \det_{\row} \left[ \Lambda \right] \right\}^{-1/2}. \label{eq:after_degen_pert}
\eea
As in the main text, the subscript ``row'' on the determinant denotes restriction to the row space of $\Lambda$. Note: in moving from \eq{eq:before_degen_pert} to \eq{eq:after_degen_pert}, we have have used degenerate perturbation theory  in the $\epsilon \to 0$ limit. 

Putting these values for $Z_\ell$ and $Z_\ell^0$ back into \eq{eq:evidence_formula}, we get
\bea
p(\data | \ell) &=& \epsilon^{\frac{\alpha-1}{2}} \frac{\sqrt{2 \pi} N^{N-\frac{\alpha}{2}}}{L^{N} \Gamma(N)} e^{-S_\ell[\phi_\ell]} \sqrt{ \frac{ \eta^{\alpha} \det_{\row} \left[\Lambda \right] }{  \det  \left[ \Lambda + \eta e^{- \phi_\ell}  \right] } }.~~~~~~~
\eea
Although both $Z_\ell$ and $Z_\ell^0$ depend strongly on the number of grid points $G$, the value for the evidence does not. The evidence does, however, depend on the regularization parameter $\epsilon$; specifically, it is proportional to $\epsilon^{\frac{\alpha-1}{2}}$. This is the basis for the claim in the main text that the evidence vanishes for $\alpha > 1$. 

In the large $\ell$ limit, $\eta \to 0$, and so 
\bea
\det  \left[ \Lambda + \eta e^{-\phi_\ell} \right]  \to \eta^{\alpha}  \det_{\ker} \left[ e^{-\phi_\infty} \right]  \det_{\row}  \left[ \Lambda \right]
\eea
where ``ker'' denotes restriction to the kernel of $\Lambda$. As a result,
\bea
p(\data | \infty) &=& \epsilon^{\frac{\alpha-1}{2}} \frac{\sqrt{2 \pi} N^{N-\frac{\alpha}{2}}}{L^{N} \Gamma(N)} \frac{e^{-S_\infty[\phi_\infty]} }{\sqrt{\det_{\ker} \left[ e^{-\phi_\infty} \right] } }.
\eea
Taking the ratio of these expressions for $p(\data | \ell)$ and $p(\data | \infty)$, we recover the formula for the evidence ratio $E$ in \eq{eq:evidence_ratio}. Note that $E$, unlike the evidence itself, does not depend on the regularization parameter $\epsilon$.

%
% Appendix: Derivation of the $K$ coefficient
%
\section{Derivation of the $K$ coefficient}

The goal of this section is to clarify the large length scale $(\ell \to \infty)$ behavior of
\bea
\ln E &=& S_\infty[\phi_\infty] - S_\ell[\phi_\ell] \\ 
& & + \frac{1}{2} \ln \left\{ \frac{  \det_{\ker} \left[ e^{-\phi_\infty} \right]  \det_{\row} \left[ \Lambda \right] }{ \eta^{-\alpha} \det  \left[ \Lambda + \eta e^{-\phi_\ell} \right] } \right\}.\nonumber
\eea
To do this we expand $\ln E$ as a power series in $\eta$ about $\eta = 0$. We will find that $\ln E = K \eta + O(\eta^2)$ where $K$ is a nontrivial coefficient, given by \eq{eq:K}, that can be either positive or negative depending on the data. 

We carry out this expansion in three steps: 
\begin{enumerate}
\item Expand $\phi_\ell$ to first order in $\eta$.
\item Expand $S_\ell[\phi_\ell]$ to first order in $\eta$.
\item Expand $\ln \det[\Lambda + \eta e^{-\phi_\ell}]$ to first order in $\eta$.
\end{enumerate}

In what follows we will use the bracket notation of Appendix B. The eigenvalues $\lambda_i$ and corresponding eigenfunctions $\psi_i(x)$ of $\Lambda$ are taken to satisfy 
\bea
\braket{\psi_i | \psi_j} &=& \delta_{ij}~~~{\rm for~all}~i,j, \\
\lambda_i &=& 0~~~{\rm for}~i \leq \alpha,
\eea
and
\bea
\braket{\psi_i | e^{-\phi_\infty} | \psi_j} = \delta_{ij} \zeta_j~~~{\rm for}~i,j \leq \alpha
\eea
for some positive numbers $\zeta_j$. We will also make use of the following indexed quantities,
\bea
v_i &=& L \braket{\psi_i | Q_\infty - R} = \int dx\, (Q_\infty - R) \psi_i \\
z_{ij} &=& L \braket{\psi_i | Q_\infty | \psi_j} = \int dx\, Q_\infty \psi_i \psi_j \\
z_{ijk} &=& L \braket{\psi_i | Q_\infty \psi_j | \psi_k} = \int dx\, Q_\infty \psi_i \psi_j \psi_k.
\eea

%%%
\subsection{Expansion of $\phi_\ell$ to first order in $\eta$.}

We begin by representing $\phi_\ell$ as a power series in $\eta$. Abusing our subscript notation somewhat, we write
\bea
\phi_\ell = \phi_\infty + \eta \phi_1 + \eta^2 \phi_2 + \cdots. \label{eq:phi_ell_expansion}
\eea
Plugging this expansion into the equation of motion,
\bea
0 = \Lambda \phi_\ell + \eta(L R - e^{-\phi_\ell}), \label{eq:appendix_eom}
\eea
then collecting terms of equal order in $\eta$, we get,
\bea
0 &=& \Lambda \phi_\infty \nonumber \\
& & + \eta \left( \Lambda \phi_1 + L R -  e^{-\phi_\infty} \right) \nonumber \\
&& + \eta^2 \left( \Lambda \phi_2 + e^{-\phi_\infty} \phi_1 \right) + \cdots \label{eq:eom_expansion}
\eea
This expansion must vanish at each order in $\eta$. At lowest order in $\eta$ we recover $\Lambda \phi_\infty = 0$, which just reflects the restriction of $\phi_\infty$ to the kernel of $\Lambda$. At first order in $\eta$,
\bea
0 = \Lambda \phi_1 + LR - e^{-\phi_\infty} \label{eq:eom_first_order},
\eea
which we will now proceed to investigate. 

To compute $\phi_1$, we first write it in terms of the eigenfunctions of $\Lambda$, i.e.,
\bea
\phi_1(x) = \sum_i A_i \psi_i(x)
\eea
for appropriate coefficients $A_i$. Taking the inner product of \eq{eq:eom_first_order} with $\bra{\psi_i}$, we get
\bea
0 &=& \lambda_i A_i +  L \braket{\psi_i | R - Q_\infty}\\
& =& \lambda_i A_i - v_i.
\eea
Since $\lambda_i > 0$ for $i > \alpha$, we find that
\bea
A_i &=& \frac{v_i}{\lambda_i},~~~i > \alpha. \label{eq:coeffs_from_first_order}
\eea

As yet we have no information about the values $A_i$ for $i \leq \alpha$. For this we need to consider the second order term in \eq{eq:eom_expansion}. Starting from
\bea
0 = \Lambda \phi_2 + e^{-\phi_\infty} \phi_1
\eea
and dotting each side with $\bra{\psi_j}$ for $j \leq \alpha$, we find that 
\bea
0 &=& \braket{\psi_j | \Lambda | \phi_2} + \braket{\phi_i | e^{-\phi_\infty} | \phi_1} \\
&=& \sum_i A_i \braket{\psi_j |e^{-\phi_\infty} | \psi_i} \\
&=& A_j \zeta_j + \sum_{i > \alpha} A_i z_{ij}
\eea
Applying \eq{eq:coeffs_from_first_order}, we thus see that
\bea
A_j &=& - \sum_{i > \alpha} \frac{v_i z_{ij} }{\lambda_i \zeta_j},~~~j \leq \alpha.
\eea
This completes our computation of the $A_i$ coefficients. We find that
\bea
\phi_\ell = \phi_\infty + \eta \left[ \sum_{i > \alpha} \frac{v_i}{\lambda_i} \psi_i - \sum_{\substack{i > \alpha \\ j \leq \alpha}}  \frac{v_i z_{ij} }{\lambda_i \zeta_j} \psi_j \right] + O(\eta^2).
\eea

%%%
\subsection{Expansion of $S_\ell[\phi_\ell]$ to first order in $\eta$.}

The action $S_\ell[\phi]$ can be expressed as
\bea
S_\ell[\phi] = N \left\{ \frac{\eta^{-1}}{2} \braket{\phi | \Lambda | \phi}  + L \braket{\phi | R} + \int \frac{dx}{L} e^{-\phi} \right\}.~~~
\eea
Using this expression together with the expansion in \eq{eq:phi_ell_expansion}, we find that the value of the action $S_\ell$ at its minimum $\phi_\ell$ is
\bea
S_\ell [ \phi_\ell]  &=& S_\infty[\phi_\infty] \nonumber \\
& & + N \eta \left\{ \frac{1}{2} \braket{\phi_1 | \Lambda | \phi_1} + L \braket{\phi_1 | R - Q_\infty} \right\}  \nonumber \\
& & + O(\eta^2) \label{eq:inner_product_terms}.
\eea
The first inner product term in \eq{eq:inner_product_terms} is
\bea
\frac{1}{2} \braket{\phi_1 | \Lambda | \phi_1} 
&=& \frac{1}{2} \sum_{i > \alpha} \frac{v_i^2}{\lambda_i},
\eea
while second is
\bea
L \braket{\phi_1 | R - Q_\infty } &=& - \sum_{i > \alpha} \frac{v_i^2}{\lambda_i}.
\eea
This gives the rather simple result, 
\bea
S_\ell[ \phi_\ell] - S_\infty[\phi_\infty] = - \eta \sum_{i > \alpha} \frac{N v_i^2}{2 \lambda_i} + O(\eta^2).\label{eq:action_expansion}
\eea

%%%
\subsection{Expansion of $\ln \det[\Lambda + \eta e^{-\phi_\ell}]$ to first order in $\eta$.}

We first outline how we will go about computing $\ln \det \Gamma$ where $\Gamma = \Lambda + \eta e^{-\phi_\ell}$. Calculating this quantity requires calculating the eigenvalues of $\Gamma$. We will use $\gamma_i$ and $\rho_i$ to denote the eigenvalues and corresponding orthonormal eigenfunctions of $\Gamma$, i.e.,
\bea
\Gamma \rho_i = \gamma_i \rho_i. \label{eq:gamma_spectrum}
\eea
and
\bea
\braket{\rho_i | \rho_j} =\delta_{ij} \label{eq:rho_normalization}.
\eea
Our primary task is to compute each eigenvalue $\gamma_i$ as a power series in $\eta$:
\bea
\gamma_i = \lambda_i + \eta \gamma_i^1 + \eta^2 \gamma_i^2 + \cdots. \label{eq:eigenvalue_expansion}
\eea
This task, as we will see, also requires computing the eigenfunctions $\rho_i$ as power series in $\eta$:
\bea
\rho_i = \psi_i + \eta \rho_i^1 + \eta^2 \rho_i^2 + \cdots. \label{eq:eigenfunction_expansion}
\eea
As usual, it will help to expand $\rho_i^1$ in the eigenfunctions of $\Lambda$. We write
\bea
\rho_i^1(x) &=& \sum_j B_j^i \psi_j(x), \label{eq:rho_expansion}
\eea
and will soon proceed to compute the coefficients $B_j^i$.

Keeping in mind that $\lambda_i > 0$ for $i > \alpha$, and $\lambda_j = 0$ for $j \leq \alpha$, we see that
%\begin{widetext}
\bea
\ln \det \Gamma &=& \ln \prod_i \gamma_i \\ 
&=& \ln \left\{ \eta^{\alpha} \left( \prod_{j \leq \alpha} \gamma_i^1 \right) \left( \prod_{i > \alpha} \lambda_i \right) \right\} \\
&& + \eta \left\{ \sum_{i > \alpha} \frac{\gamma_i^1}{\lambda_i} + \sum_{i \leq \alpha} \frac{\gamma_i^2}{\gamma_i^1} \right\}  + O(\eta^2). \label{eq:plug_into_this}
\eea
%\end{widetext}
So while the larger eigenvalues of $\Gamma$ need only be computed to first order in $\eta$, the $\alpha$ lowest eigenvalues of $\Gamma$ must actually be computed to second order in $\eta$. Performing this second order calculation will require that we (partially) compute the eigenfunctions $\rho_i$ to first order in $\eta$, i.e.\ compute (some of) the coefficients $B_j^i$ in \eq{eq:rho_expansion}.

Plugging \eq{eq:eigenvalue_expansion} and \eq{eq:eigenfunction_expansion} into \eq{eq:gamma_spectrum} and collecting terms by order in $\eta$, we get
\bea
0 &=& \Gamma \rho_i - \gamma_i \rho_i \\
&=& (\Lambda + \eta e^{-\phi_\infty} + \eta^2 e^{-\phi_\infty} \phi_1) (\psi_i + \eta \rho_i^1 + \eta^2 \rho_i^2) \nonumber \\
& & -(\lambda_i + \eta \gamma_i^1 + \eta^2 \gamma_i^2)(\psi_i + \eta \rho_i^1 + \eta^2 \rho_i^2) + O(\eta^3) \\
&=&  \left(\Lambda \psi_i - \lambda_i \psi_i  \right) \nonumber \\
&& + \eta \left( \Lambda \rho_i^1 + e^{-\phi_\infty} \psi_i  - \lambda_i \rho_i^1 - \gamma_i^1 \psi_i \right)  \nonumber \\
& & + \eta^2 \left( \Lambda \rho_i^2 + e^{-\phi_\infty} \rho_i^1 - e^{-\phi_\infty} \phi_1 \psi_i - \lambda_i \rho_i^2 - \gamma_i^1 \rho_i^1 - \gamma_i^2 \psi_i \right) \nonumber \\ 
& &  + O(\eta^3). \label{eq:gamma_spectrum_expansion} 
\eea
The coefficient of each term in this expansion must vanish. 
From the zeroth order term of \eq{eq:gamma_spectrum_expansion} we recover the eigenvalue equation $\Lambda \psi_i = \lambda_i \psi_i$, which we already knew. From the first order term we get
\bea
0 = \Lambda \rho_i^1 + e^{-\phi_\infty} \psi_i  - \lambda_i \rho_i^1 - \gamma_i^1 \psi_i.
\eea
Dotting this with $\bra{\psi_k}$ and using \eq{eq:rho_expansion} then gives
\bea
0 &=& \braket{\psi_k | \Lambda | \rho_i^1} + \braket{\psi_k | e^{-\phi_\infty} | \psi_i} - \lambda_i \braket{\psi_k | \rho_i^1} - \gamma_i^1 \braket{\psi_k | \psi_i} ~~~~~~~~\\
&=& (\lambda_k - \lambda_i) B_k^i + z_{ik} -  \gamma_i^1 \delta_{ik}. \label{eq:first_order_equation}
\eea
If we set $k = i$, we recover the standard first order correction to the eigenvalues, namely
\bea
\gamma_i^1 = z_{ii}~~~{\rm for~all}~~~i,
\eea
in particular, 
\bea
\gamma_j^1 = \zeta_j~~~{\rm for}~~~{j \leq \alpha}.
\eea
We also see by inspection of \eq{eq:first_order_equation} that
\bea
B_k^i = -\frac{z_{ik}}{\lambda_k} ~~~{\rm for}~~~{i \leq \alpha,~ k > \alpha}. \label{eq:B_coeffs}
\eea
Moreover, from the normalization requirement of \eq{eq:rho_normalization},
\bea
1 &=& \braket{\rho_i | \rho_i} \\
&=& \braket{\psi_i | \psi_i} + 2 \eta \braket{\psi_i | \rho_i} + O(\eta^2) \\
&=& 1 + 2 \eta B_i^i + O(\eta^2),
\eea
from which we conclude that 
\bea
B_i^i = 0~~~{\rm for~all}~~~i.
\eea

Turning to the second-order term in \eq{eq:gamma_spectrum_expansion}, we now consider the requirement
\bea
0 = \Lambda \rho_i^2 + e^{-\phi_\infty} \rho_i^1 - e^{-\phi_\infty} \phi_1 \psi_i - \lambda_i \rho_i^2 - \gamma_i^1 \rho_i^1 - \gamma_i^2 \psi_i.~~~~~~~~
\eea
Choosing $i \leq \alpha$, dotting with $\bra{\psi_i}$, and using the fact that $\lambda_i = 0$, we find that
\bea
0 &=&\braket{ \psi_i | e^{-\phi_\infty} | \rho_i^1} - \braket{ \psi_i | e^{-\phi_\infty} \phi_1 |\psi_i} - \gamma_i^1 \braket{\psi_i | \rho_i^1} -  \gamma_i^2~~~~~~~ \\
&=& \sum_j B_j^i z_{ij} - \sum_j A_j z_{iij} - \gamma_i^1 B_i^i - \gamma_i^2 \label{eq:analyze_this}
\eea
Now consider the first term of \eq{eq:analyze_this}. Because $B_i^i = 0$ and $z_{ij} = \zeta_i \delta_{ij}$ for $i,j \leq \alpha$, no $j \leq \alpha$ terms  contribute to this sum. The third term also vanishes due to $B_i^i = 0$. So for $i \leq \alpha$,
\bea
\gamma_i^2 &=& \sum_{j > \alpha} B_j^i z_{ij} - \sum_{j > \alpha} A_j z_{iij}- \sum_{j \leq \alpha} A_j z_{iij} \\
&=& 
- \sum_{j > \alpha} \frac{z_{ij}^2}{\lambda_j} 
- \sum_{j > \alpha} \frac{v_j z_{iij}}{\lambda_j} 
+ \sum_{\substack{j \leq \alpha \\ k > \alpha}} \frac{v_k z_{jk} z_{iij}}{\lambda_k \zeta_j} 
\eea

Having computed $\gamma_i^1$ for all $i$ and $\gamma_i^2$ for $i \leq \alpha$, we can now find $\ln \det \Gamma$. Plugging in our results for $\gamma_i^1$ and $\gamma_i^2$ and using
\bea
\prod_{j \leq \alpha}  \zeta_j  =  \det_{\ker}[ e^{-\phi_\infty}],~~~~~\prod_{i > \alpha} \lambda_i = \det_{\row} [ \Lambda ],
\eea
we get what we set out to find:
\bea
\ln \det \Gamma  &=& \ln \left\{ \eta^{\alpha} \det_{\ker}[ e^{-\phi_\infty}] \det_{\row} [ \Lambda ] \right\} \nonumber \\
&& + \eta \left\{ 
\sum_{i > \alpha} \frac{z_{ii}}{\lambda_i} 
- \sum_{\substack{j > \alpha \\ i \leq \alpha}} \frac{z_{ij}^2 + v_j z_{iij}}{\lambda_j \zeta_i}  
+ \sum_{\substack{k > \alpha \\ i, j \leq \alpha}} \frac{v_k z_{jk} z_{iij}}{\lambda_k \zeta_i \zeta_j}  
\right\} \nonumber \\
& & + O(\eta^2).
 \label{eq:det_expansion}
\eea

\subsection{Putting it all together}

Putting together our results from \eq{eq:action_expansion} and \eq{eq:det_expansion}, we find that to first order in $\eta$,
\bea
\ln E &=&  S_\infty[\phi_\infty] - S_{\ell}[\phi_\ell]    \nonumber \\
& & - \frac{1}{2} \ln \left\{ \frac{\det \Gamma }{\eta^{\alpha} \det_{\ker}[ e^{-\phi_\infty}] \det_{\row} [ \Lambda ] } \right\} \\
&=& \eta \sum_{i > \alpha} \frac{N v_i^2}{2 \lambda_i} \nonumber \\
&& - \frac{\eta}{2} \left\{ 
\sum_{i > \alpha} \frac{z_{ii}}{\lambda_i} 
- \sum_{\substack{j > \alpha \\ i \leq \alpha}} \frac{z_{ij}^2 + v_j z_{iij}}{\lambda_j \zeta_i}  
+ \sum_{\substack{k > \alpha \\ i, j \leq \alpha}} \frac{v_k z_{jk} z_{iij}}{\lambda_k \zeta_i \zeta_j}  
\right\} ~~~~~~~\\
&=& K \eta,
\eea
where 
\bea
K = 
\sum_{i > \alpha} \frac{N v_i^2 - z_{ii}}{2 \lambda_i} 
+ \sum_{\substack{j > \alpha \\ i \leq \alpha}} \frac{z_{ij}^2 + v_j z_{iij}}{2 \lambda_j \zeta_i}  
- \sum_{\substack{k > \alpha \\ i, j \leq \alpha}} \frac{v_k z_{jk} z_{iij}}{2 \lambda_k \zeta_i \zeta_j}. ~~~~~~~~
\eea
The formula for $K$ in \eq{eq:K} is obtained by renaming the indices $i,j,k$ in the second and third terms.

%
% Appendix D: Computing the MAP path
%
\section{Predictor-corrector homotopy algorithm \label{sec:algorithm}}

%%%
\subsection{Computing the MaxEnt density}

We saw in the main text that adopting the prior defined by the action in \eq{eq:prior} renders $\phi_\infty$ a polynomial of order $\alpha-1$, i.e.,
\bea
\phi_\infty(x) = \sum_{i=0}^{\alpha-1} a_i x^i \label{eq:maxentcoeffs}
\eea
for some vector of coefficients $\vec{a} = (a_0, a_1, \ldots, a_{\alpha-1})$. The problem of computing the MaxEnt density $Q_\infty$ therefore reduces to finding the vector $\vec{a}$ that minimizes the posterior action
\bea
S_\infty(\vec{a}) &=& N \int \frac{dx}{L} \left\{ L R \sum_{i=0}^{\alpha-1} a_i x^i + \exp \left[- \sum_{i=0}^{\alpha-1} a_i x^i \right] \right\}. ~~~~\label{eq:actionasfunctionofcoeefs}
\eea

Following Ormoneit and White \cite{Ormoneit:1999}, we solve this optimization problem using the Newton-Raphson algorithm with backtracking. Starting at $\vec{a}^0 = \vec{0}$, we iterate
\bea
\vec{a}^n \to \vec{a}^{n+1} = \vec{a}^n + \gamma_n \delta \vec{a}^n \label{eq:maxentstep}
\eea
where the vector $\delta \vec{a}^n$ is the solution to
\bea
\sum_{j=0}^{\alpha-1} \left. \frac{\partial^2 S}{\partial a_i \partial a_j} \right|_{\vec{a}^n} \delta a^n_j = - \left. \frac{\partial S}{\partial a_i}\right|_{\vec{a}^n} \label{eq:maxent_equation}
\eea
and $\gamma_n$ is some number in the interval $(0,1]$. From \eq{eq:actionasfunctionofcoeefs}, 
\bea
\frac{\partial S}{\partial a_i} &=& N \mu_i - N \int \frac{dx}{L}\, x^i \exp \left[- \sum_{k=1}^{\alpha-1} a_k x^k \right],
\eea
where $\mu_i = \int dx\, R\, x^i$ is the $i$'th moment of the data, and
\bea
\frac{\partial^2 S}{\partial a_i \partial a_j} &=& N \int \frac{dx}{L}\, x^{i+j} \exp \left[- \sum_{k=1}^{\alpha-1} a_k x^k \right].
\eea
The Hessian matrix $H$, with elements $H_{ij} = \frac{\partial^2 S}{\partial a_i \partial a_j}$, is positive definite at all $\vec{a}$. This is readily seen from the fact that for any vector $\vec{w}$,
\bea
\vec{w}^\top H \vec{w} = N \int \frac{dx}{L} \left(\sum_i x^i w_i \right)^2 e^{-\sum_k a_k x^k} > 0.
\eea
\eq{eq:maxent_equation} will therefore always yield a unique solution for $\delta \vec{a}^n$. 

The scalar $\gamma_n$ is chosen so that the change in the action in each iteration is commensurate with the linear approximation. Specifically, $\gamma_n$ is first set to $1$. Then, if 
\bea
S_\infty(\vec{a}^n + \gamma_n \delta \vec{a}^n) - S_\infty(\vec{a}^n)  < \frac{\gamma_n}{2} \sum_{i = 0}^{\alpha-1} \left. \frac{\partial S}{\partial a_i} \right|_{\vec{a}^n} \delta a^n_i ~~~\label{eq:gamma_condition}
\eea
is not satisfied, $\gamma_n$ is reduced by factors of $\frac{1}{2}$ until \eq{eq:gamma_condition} holds. This ``dampening'' of the Newton-Raphson method is sufficient to guarantee convergence \cite{Ormoneit:1999, Boyd:2009}. The algorithm is terminated when the magnitude of the change in the action, $|S_\infty(\vec{a}^{n+1}) - S_\infty(\vec{a}^n)|$, falls below a specified tolerance. 

%%%
\subsection{Predictor step}

The predictor step computes a change $\ell \to \ell'$ in the length scale, as well as an approximation to the corresponding change $\phi_{\ell} \to \phi_{\ell'}$ in the MAP field. Specifically, the predictor step returns a scalar $\delta t$ and a function $\rho(x)$ such that,
\bea
t' = t + \delta t
\eea
and
\bea
\phi_{\ell'}(x) \approx \phi_{\ell}^{(0)}(x) = \phi_\ell(x) + \rho(x) \delta t,
\eea
where $t = \ln \eta$ is a numerically convenient reparametrization of $\ell$. To determine the function $\rho$, we examine the equation of motion, \eq{eq:appendix_eom}, at $\ell'$:
\bea
0 &=& \Lambda \phi_{\ell'} + \eta'(L R - e^{-\phi_{\ell'}}) \\
&=& \Lambda (\phi_\ell + \rho \delta t) + \eta e^{\delta t} (L R - e^{-(\phi_\ell + \rho \delta t)}) \\
&=& \Lambda \phi_\ell + \eta (L R - e^{-\phi_\ell})  \\
& & + \delta t \left\{ \left[ \Lambda + \eta e^{-\phi_\ell} \right] \rho + \eta (L R - e^{-\phi_\ell} ) \right\} + O(\delta t^2). \nonumber
\eea
The $O(1)$ term vanishes due to $\phi^(n)$ satisfying the equation of motion at $\ell$. The $O(\delta t)$ term must therefore vanish as well. We thus obtain a linear equation,
\bea
\left[ \Lambda + \eta e^{-\phi_\ell} \right] \rho = \eta (e^{-\phi_\ell} - L R),
\eea
which can be numerically solved for $\rho$.
The scalar $\delta t$ is then chosen to satisfy the distance criterion,
\bea
\epsilon^2 &=& D_{\rm geo}^2(Q_\ell, Q_{\ell'}) \\
&\approx&  \int dx \frac{(Q_\ell - Q_{\ell'})^2}{Q_\ell} \\
&\approx& (\delta t)^2 \int dx\, Q_\ell\, \rho^2.
\eea
We therefore set
\bea
\delta t = \pm \frac{\epsilon}{\sqrt{\int dx\, Q_\ell\, \rho^2}},
\eea
with the sign of $\delta t$ chosen according to the direction we wish to traverse the MAP curve. 

%%%
\subsection{Corrector step}

The purpose of the corrector step is to accurately solve the equation of motion, \eq{eq:appendix_eom}, at fixed length scale $\ell$. This step is initially used to compute $Q_{\ell_0}$ at the starting length scale $\ell_0$. It is then employed to hone in on the MAP density at each new length scale chosen by the predictor step of the homotopy algorithm.

As with the computation of the MaxEnt density, this corrector step uses the Newton-Raphson algorithm with backtracking. Starting from a hypothesized field $\phi^{(0)}$ (e.g., returned by the predictor step), the iteration
\bea
\phi^{(n)} \to \phi^{(n+1)} = \phi^{(n)} + \gamma_n \delta \phi^{(n)} \label{eq:another_iteration}
\eea
is performed. The function $\delta \phi^{(n)}$ and scalar $\gamma_n$ are chosen so that this iteration converges to the true field $\phi_\ell$. To derive the field perturbation $\delta \phi^{(n)}$, we provisionally set $\gamma_n = 1$ and plug the above formula for $\phi^{(n+1)}$ into the equation of motion, \eq{eq:appendix_eom}. Keeping only terms of order $\delta \phi^{(n)}$ or less, we see that the field perturbation $\delta \phi^{(n)}$ is the solution to the linear equation, 
\bea
\left[ \Lambda + \eta e^{-\phi^{(n)}} \right] \delta \phi^{(n)} = \eta ( e^{-\phi} - LR) - \Lambda \phi^{(n)},
\eea
which we solve numerically. As before, $\gamma_n$ is then chosen so that so that the action decreases by an amount commensurate with the linear approximation, i.e.,
\bea
S_\ell[\phi^{(n+1)}] - S_\ell[\phi^{(n)}]  <   \frac{\gamma_n}{2} \int dx \left. \frac{\delta S_\ell}{\delta \phi(x)} \right|_{\phi^{(n)}} \delta \phi^{(n)}.~~~~~~
\eea
This iterative process is terminated when the magnitude of the change in the action, $|S_\ell[\phi^{(n+1)}] - S_\ell[\phi^{(n)}]|$, falls below a specified tolerance. 

%
% Discussion of Silverman
%
\section{Maximum penalized likelihood and Bayesian field theory}

In statistics there is a class of nonparametric techniques for estimating smooth functions called ``maximum penalized likelihood'' estimation \cite{Silverman:1982,Eggermont:2001,Gu:2013}. The central idea behind these methods is to maximize the likelihood function modified by a heuristic roughness penalty. In this context, Silverman (1982) proposed using $-S_\ell[\phi]$, defined in \eq{eq:action}, as the penalized likelihood function for probability density estimation \cite{Silverman:1982}. This choice was motivated by the observation that, when $\ell = \infty$, one recovers a moment-matching distribution having a familiar parametric form. This early work by Silverman is therefore relevant to the results reported here. 

However, the results reported here move beyond \cite{Silverman:1982} in a number of critical ways. First, the connection with MaxEnt estimation was not recognized in \cite{Silverman:1982}, nor was the fact that the MAP density $Q_\ell$ matches the same moments as $Q_\infty$ even at finite values of $\ell$. Moreover, periodic boundary conditions on $Q_\ell$ were assumed in much of the analysis described in \cite{Silverman:1982}, and the contradiction between these boundary conditions and the results for $\ell = \infty$ was not discussed. 

Perhaps most importantly, the shortcomings of the maximum penalized likelihood approach highlight the importance of adopting an explicit Bayesian interpretation. Although the penalized likelihood context of \cite{Silverman:1982} and later work (see \cite{Eggermont:2001}) was sufficient to motivate the formula for $S_\ell[\phi]$, it provided no motivation for computing the evidence $p(\data | \ell)$. Without the Bayesian notion of evidence, it is unclear how to determine the optimal smoothness length scale $\ell^*$ without resorting to empirical methods, such as cross-validation. By contrast, the Bayesian interpretation introduced by Bialek et al.\ \cite{Bialek:1996} and built upon here transparently motivates the computation of $p(\data | \ell)$, thereby providing an explicit  criterion for choosing $\ell^*$. In particular, this Bayesian interpretation is essential for our derivation of the $K$ coefficient in \eq{eq:K} of the main text. 

%
% References
%
\bibliography{14_maxent}

\end{document}